\documentclass[10pt,journal,compsoc,twoside]{IEEEtran}

\usepackage{amsmath,amssymb,amsfonts}
\usepackage{algorithmic}
\usepackage{graphicx}
\usepackage{textcomp}
\usepackage{xcolor}
\usepackage[nocompress]{cite}
\usepackage[pass]{geometry}
\usepackage{fancyhdr}
\usepackage[normalem]{ulem}
\usepackage[hyphens]{url}
\usepackage{hyperref}
\usepackage{color}
\usepackage{soul}
\usepackage{diagbox}
\usepackage{enumitem}
\usepackage{relsize}
\usepackage[square, numbers]{natbib}
\usepackage[labelfont={bf,small},textfont={bf,small}]{caption}
\usepackage{multirow}
\usepackage{rotating}
\usepackage{array}

\def\BibTeX{{\rm B\kern-.05em{\sc i\kern-.025em b}\kern-.08em
    T\kern-.1667em\lower.7ex\hbox{E}\kern-.125emX}}

\newcommand{\ignore}[1]{}

\newcommand{\SSQ}{S{\kern-0.2em\raise0.8ex\hbox{\smaller[2] 2}}\ }
\newcommand\upcite[1]{\ \cite{#1}}
\newcommand\cref[1]{\autoref{#1}}
\newcommand\Cref[1]{\autoref{#1}}
\renewcommand \thetable{\Roman{table}}

\usepackage{enumitem}
\setenumerate[1]{itemsep=5pt,partopsep=0pt,parsep=\parskip,topsep=5pt}
\setitemize[1]{itemsep=5pt,partopsep=0pt,parsep=\parskip,topsep=5pt}
\setdescription{itemsep=5pt,partopsep=0pt,parsep=\parskip,topsep=5pt}
\usepackage{booktabs}
\usepackage[flushleft]{threeparttable}
\usepackage{bigstrut}

\title{S\raise0.7ex\hbox{\Large 2} Engine: A Novel Systolic Architecture for Sparse Convolutional Neural Networks}
\author{Jianlei~Yang,~\IEEEmembership{Senior~Member,~IEEE,}
        Wenzhi~Fu,
        Xingzhou~Cheng,
        Xucheng~Ye,
        Pengcheng~Dai,
        and~Weisheng~Zhao,~\IEEEmembership{Fellow,~IEEE}

\IEEEcompsocitemizethanks{
\IEEEcompsocthanksitem This work is supported by National Natural Science Foundation of China (Grant No. 62072019, 61602022).
\IEEEcompsocthanksitem Corresponding authors are Jianlei Yang and Weisheng Zhao, Email: \url{jianlei@buaa.edu.cn}, \url{weisheng.zhao@buaa.edu.cn}.
\IEEEcompsocthanksitem J. Yang, W. Fu, X. Cheng and X. Ye are with School of Computer Science and Engineering, Beihang University, Beijing, China.
\IEEEcompsocthanksitem P. Dai and W. Zhao are with School of Integrated Circuits and Engineering, Beihang University, Beijing, China.
}

\thanks{The source code of this paper is publicly available on: \newline
\indent\indent \url{https://github.com/BUAA-CI-Lab/S2EngineCompiler} \newline
\indent\indent \url{https://github.com/BUAA-CI-Lab/S2EngineSimulator}.}
}
\begin{document}

\pagestyle{plain}

\IEEEtitleabstractindextext{%
\begin{abstract}

Convolutional neural networks (CNNs) have achieved great success in performing cognitive tasks. 
However, execution of CNNs requires a large amount of computing resources and generates heavy memory traffic, which imposes a severe challenge on computing system design.
Through optimizing parallel executions and data reuse in convolution, systolic architecture demonstrates great advantages in accelerating CNN computations. 
However, regular internal data transmission path in traditional systolic architecture prevents the systolic architecture from completely leveraging the benefits introduced by neural network sparsity. 
Deployment of fine-grained sparsity on the existing systolic architectures is greatly hindered by the incurred computational overheads. In this work, we propose \SSQ Engine -- a novel systolic architecture that can fully exploit the sparsity in CNNs with maximized data reuse.
\SSQ Engine transmits compressed data internally and allows each processing element to dynamically select an aligned data from the compressed dataflow in convolution. 
Compared to the na\"ive systolic array, \SSQ Engine achieves about $3.2\times$ and about $3.0\times$ improvements on speed and energy efficiency, respectively. 

\end{abstract}
}
\maketitle
\IEEEdisplaynontitleabstractindextext
\IEEEpeerreviewmaketitle

\IEEEraisesectionheading{\section{Introduction}\label{sec1}}

\IEEEPARstart{C}{onvolutional} neural networks (CNNs) have made remarkable success in modern artificial intelligence (AI) applications \cite{redmon2016you}. 
The required training data size and model complexity, however, keep increasing for better performance in a large variety of applications.
The incurred high computational cost and data movement bandwidth are hardly supported by conventional computing platforms and hence, motivating recent huge investment on the corresponding accelerators \cite{nvdla}. 
Among these designs, systolic architecture \cite{7-hwang1989systolic} -- a specialized processing element network designed for massive parallelization and extensive data reuse, has been proved efficient in performing CNN computations.
In addition to academic research~\cite{10-wei2017automated, kung2018packing}, systolic architectures were also adopted in some industrial practices of deep neural network (DNN) accelerators \cite{11-jouppi2017datacenter,tpuv2,Norman2018a,xilinx2018wp}.

The highly regularized layout and internal data movement path of systolic arrays make engineering realization of the design very efficient. However, such a regularity also prevents exploiting irregular computation patterns that frequently appear in sparse CNNs. 
Sparsity of deep CNNs has been proven important to minimize computation workloads and model size \cite{wen2016learning, 18-han2015learning, mao2017exploring}.
State-of-the-art pruning algorithms can reduce the model size by $>10\times$ \cite{19-han2015deep} and computational cost by $>4\times$ \cite{18-han2015learning} with negligible accuracy loss. 
However, due to the large variety of the irregularities, the sparsity is not fully exploited by the existing accelerators and introduces significant design overheads. 
For example, Cambricon-X only considers the weight sparsity \cite{15-20-zhang2016cambricon} while Cnvlutin only deploys the feature sparsity \cite{21-albericio2016cnvlutin}.
Cambricon-S can fully deploy both the weight and feature sparsity but requires the sparsity pattern to be coarse-grained, which greatly limits the application scope \cite{22-zidong2018cambricon-s}.
SCNN supports fine-grained sparsity in both feature and weight but introduces significant computational overhead due to the required additional coordinates transformation \cite{17-parashar2017scnn}.
SparTen \cite{gondimalla2019sparten} also utilizes both feature and weight sparsity, but the energy efficiency is significantly degraded due to the required additional logic for inner-join operations.

In this work, we propose \textbf{\SSQ} Engine -- a novel \textbf{S}ystolic architecture for \textbf{S}parse convolutional neural networks. 
Different from the existing systolic approaches \cite{10-wei2017automated, 11-jouppi2017datacenter}, compressed feature and weight flows are fed into the systolic array and the aligned pairs can be dynamically selected from the compressed dataflow by each processing element (PE).
Our approach solves the contradiction between the regularity of data transmission and the irregularity of sparsity such that the sparsity in CNNs can be fully exploited.
Data reuse could be efficiently implemented in \SSQ Engine by introducing an associated collective element (CE) array for the PE array to reduce external buffer access. 
Furthermore, fine-grained mixed-precision data processing is also supported by \SSQ Engine to satisfy the varying precision requirement of CNNs even within the same kernel \cite{3-park2018value}. 
Experimental results show that \SSQ Engine can achieve about $3.2\times$ speedup and about $3.0\times$ energy efficiency improvement compared with the existing systolic approaches.

The main contributions of our work are:
\begin{itemize}
    \item We propose a novel systolic architecture, namely, \SSQ Engine. By allowing each PE to select the aligned data pairs dynamically from the compressed dataflows, \SSQ Engine could fully exploit the sparsity during the execution of CNNs with low overhead.
    \item We introduce a collective element (CE) array to further allow data reuse during convolution procedures. This technique also works for native systolic array.
    \item \SSQ Engine solves the contradiction between the regularity of data transmission and the irregularity of sparsity, showing good robustness for different sparsity degrees.
\end{itemize}

The rest of the paper is organized as follows. \cref{sec2} provides a preliminary on CNN sparsity and a brief introduction of systolic architecture for CNN accelerators. \cref{sec3} discusses the motivation of our design by considering data reuse manner and sparsity irregularities in CNNs. \cref{sec4} illustrates the detailed architectures of proposed \SSQ Engine. \cref{sec5} explains the experimental methodology and \cref{sec6} presents the experimental results. Several related works are discussed in \Cref{sec7}, and concluding remarks are given in \Cref{sec8}.

\section{Preliminaries}\label{sec2}
In this section, we give preliminaries of both CNN and systolic architecture.

\subsection{CNN Sparsity and Quantization}

\begin{figure}[t]
    \centering
    \includegraphics[width=\columnwidth]{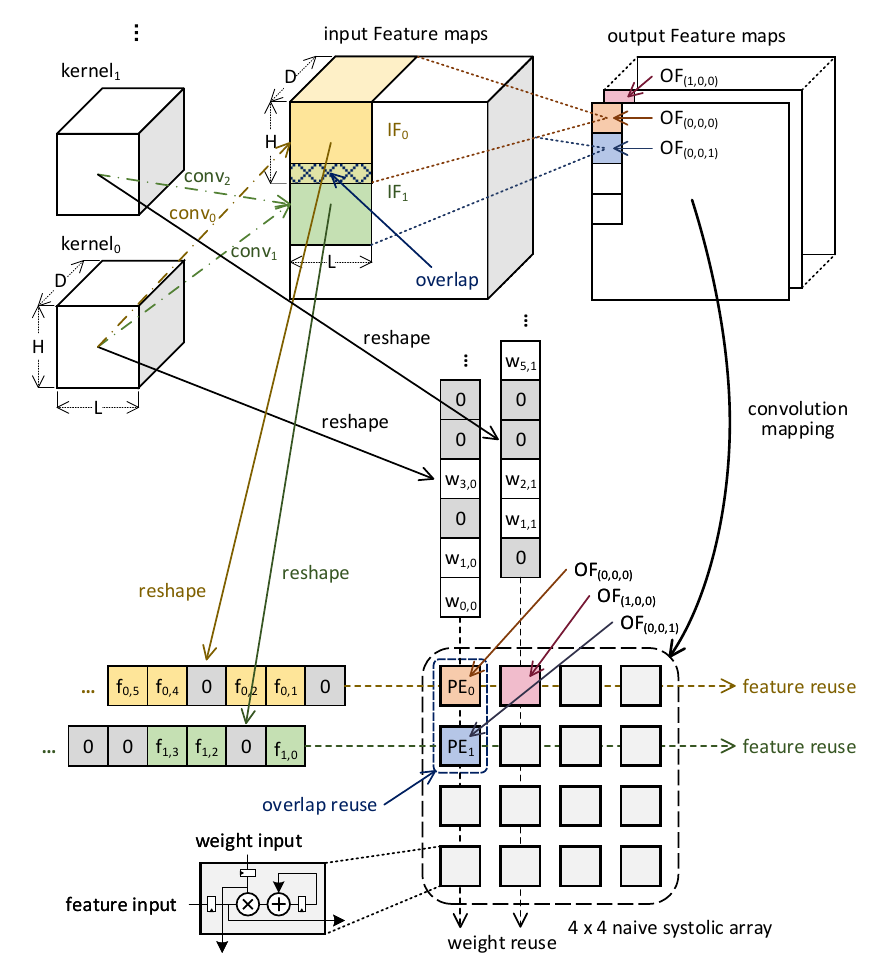}
    \caption{Illustration of data reuse manners among convolutions.}
    \label{convnet}
\end{figure}

The main computation task of CNN algorithms is performing convolution operations layer by layer. As illustrated in \cref{convnet}, convolution procedure is carried out between different kernels and input feature maps. Take the convolution of ${conv}_0$ as an example, the dimensionality of ${kernel}_0$ and ${IF}_0$ (input feature) could be represented as ${{H}\times{L}\times{D}}$, and ${conv}_0$ is defined as:
\begin{equation}
    \sum\nolimits_{i = 0}^{H-1} {\sum\nolimits_{j = 0}^{L-1} {\sum\nolimits_{k = 0}^{D-1} {{kernel}_0\left[ i \right]\left[ j \right]\left[ k \right] \times {IF}_0\left[ i \right]\left[ j \right]\left[ k \right]} } }
\end{equation}
where $i \in \left[ {0,H-1} \right]$, $j \in \left[ {0,L-1} \right]$ and $k \in \left[ {0,D-1} \right]$ represent the indices of three dimensions, respectively. After the convolution operations, nonlinear activation function (such as rectified linear unit, ReLU) is usually applied to obtain the output feature ${OF}_{(i',j',k')}$, while $i' \in \left[ {0,H'-1} \right]$, $j' \in \left[ {0,L'-1} \right]$ and $k' \in \left[ {0,D'-1} \right]$ represent the indices of three dimensions ${{H'}\times{L'}\times{D'}}$ for output feature maps, respectively.

In deep CNNs, sparsity exists both in weights and features due to the inherent redundancy of the networks. 
Weights sparsity denotes the zeros in convolution kernels and is often obtained by various pruning algorithms. 
Feature sparsity denotes the zeros in feature maps and is caused by zero activation functions. 
Sparsity levels of weights and features are defined by the percentages of the zeros existing in kernels and feature maps, respectively. 
With negligible loss on accuracy, the state-of-the-art pruning algorithms \cite{19-han2015deep, 18-han2015learning} can significantly increase the sparsity level of CNNs and lead to reduction in both model size and computation workloads.
Therefore, the sparsity of CNNs greatly affects the computing efficiency of the relevant neural network accelerators.
However, due to unstructured pruning and the randomness of input data, the sparsity is irregular and unpredictable, which is difficult for accelerators to leverage.
    
Another strategy to reduce network redundancy is called quantization, that is, using low-precision fixed-point data representation instead of using high-precision float-point one to perform inference \cite{19-han2015deep, zhou2016dorefa, Hubara2016Quantized}.
This strategy can also significantly reduce the model footprint, memory access and computational overhead of the CNN while maintaining its accuracy.
Many recent designs of neural network accelerators and GPUs already support $8$/$16$-bit fixed-point data in CNN inference.
In addition to the one-precision-fits-all approaches, mixed-precision quantization algorithms have been also developed by assigning different precision to different layers of CNNs according to the different sensitivities of each layer \cite{2-judd2015reduced}.
A fine-granularity quantization approach is also proposed by~\cite{3-park2018value}, where most data is represented with low precision (i.e., $4$-bit) while only a small portion of the data (i.e., $3\%$) is represented with high precision (i.e., $16$-bit).
These mixed-precision approaches further reduce the CNN computation cost and memory consumption.

\subsection{Systolic Architecture}

Systolic array is a specialized network of homogeneous PEs that is designed for massive parallel computing in a special-purpose system. 
Both the structure of PEs and the communication in the systolic array keep simple and regular, offering great convenience to practical implementations. 
In a typical systolic array design, all the internal PEs can get their input data from the neighboring PEs and do not need to access the external memory. 
Therefore, systolic array becomes an efficient dataflow-driven architecture and can achieve high throughput with relatively low memory bandwidth.

Several previous works adopt systolic array for CNNs acceleration with different dataflows. 
According to the taxonomy in \cite{1-chen2017using} and \cite{samajdar2018scale}, the dataflow adopted by the accelerator in \cite{10-wei2017automated} belongs to output stationary. 
Under this dataflow, as demonstrated in \cref{convnet}, each PE undertakes the computation of a convolution. 
The convolution in the same channel are allocated to the PEs in the same row. 
Input features and kernels are both reshaped into one-dimensional vectors and then fed into each row and column of the systolic array, respectively.
TPU \cite{11-jouppi2017datacenter} adopts weight stationary dataflow. 
The feature of each convolutional layer is loaded into the systolic array before the weight is fed into the array. 
Different from \cite{10-wei2017automated}, each PE in TPU does not complete an entire convolution. 
Instead, the partial accumulation is transported between adjacent PEs and the convolution computation is completed by the accumulation in the PEs along the transmission path.
Despite the difference in the dataflow, all of those designs retain the basic characteristics of the systolic array and obtain significant advantages of both speed and energy efficiency.

\section{Motivation} \label{sec3}

The efficiency of CNN accelerators is largely determined by how to exploit the network sparsity and data reuse.
In systolic architecture, however, the regular structure greatly limits its capability of supporting the network sparsity and data reuse.
In this section, we analyze the potential optimization space of systolic architecture from the perspectives of \textbf{data reuse} and \textbf{sparsity}.

\subsection{Optimization with Data Reuse}
During the computation of deep CNNs, each parameter may be accessed for many times by MAC operations, as depicted by the statistics in \cref{tab1}. 
Therefore, repeatedly loading these data from a separate memory (e.g., a DRAM or a global buffer) introduces excessive memory accesses. 
Since the energy consumption of memory accesses can be much larger than that of normal logic computations\upcite{4-hameed2010understanding, 5-horowitz20141},
reducing memory accesses, e.g., by improving data reuse, can substantially reduce the energy consumption of DNN accelerators.

\begin{table}[ht]
 \newcommand{\tabincell}[2]{\begin{tabular}{@{}#1@{}}#2\end{tabular}}
  \centering
  \begin{threeparttable}
  \caption{Average accesses per parameter by MACs in various CNNs.}
  \label{tab1}
  \small
  \begin{tabular}{r|ccc}
    \specialrule{0.8pt}{0pt}{0pt}
                  & AlexNet & VGG16  & ResNet50    \bigstrut \\
                  & \cite{9-krizhevsky2012imagenet} 
                  & \cite{simonyan2014very}  
                  & \cite{he2016deep}              \bigstrut \\ \hline
    Total MACs    & 666M    & 15.3G  & 3.86G       \bigstrut \\ \hline
    Parameters    & 2.33M   & 14.7M  & 23.5M       \bigstrut \\ \hline
    Avg. Usage of Param. & 572 & 2082 & 336 \bigstrut \\
    \specialrule{0.8pt}{0pt}{0pt}
  \end{tabular}
  \end{threeparttable}
\end{table}

We can divide the data reuse strategies into three types: \textbf{weight reuse}, \textbf{feature reuse} and \textbf{overlap reuse}.
As illustrated in \cref{convnet}, weight reuse is defined as performing convolutions between the same kernel (weights) and different input feature maps, e.g., ${conv}_0$ and ${conv}_1$ sharing the same ${kernel}_0$.
Correspondingly, feature reuse denotes convolutions between different kernels (weights) and the same feature, such as ${conv}_1$ and ${conv}_2$ sharing the same \texttt{IF}$_1$. 
Overlap reuse also exists between ${conv}_0$ and ${conv}_1$, which denotes convolutions between the same kernel (weights) and the overlapped input feature maps.

As illustrated in \cref{convnet}, the dataflow in na\"ive systolic array can naturally utilize both the weight and feature reuses by fetching data from adjacent PEs instead of external buffers. 
However, it does not support the overlap reuse which needs to access the adjacent rows of PE array. 
Therefore, the insufficient exploration of overlap reuse leads to excessive accesses on external buffers and extra buffer capacity.
As a supplement to the na\"ive design, the proposed \SSQ Engine can exploit overlap reuse and hence, support all three types of data reuse strategies.

\begin{figure}[t]
    \centering
    \includegraphics[width=0.9\columnwidth]{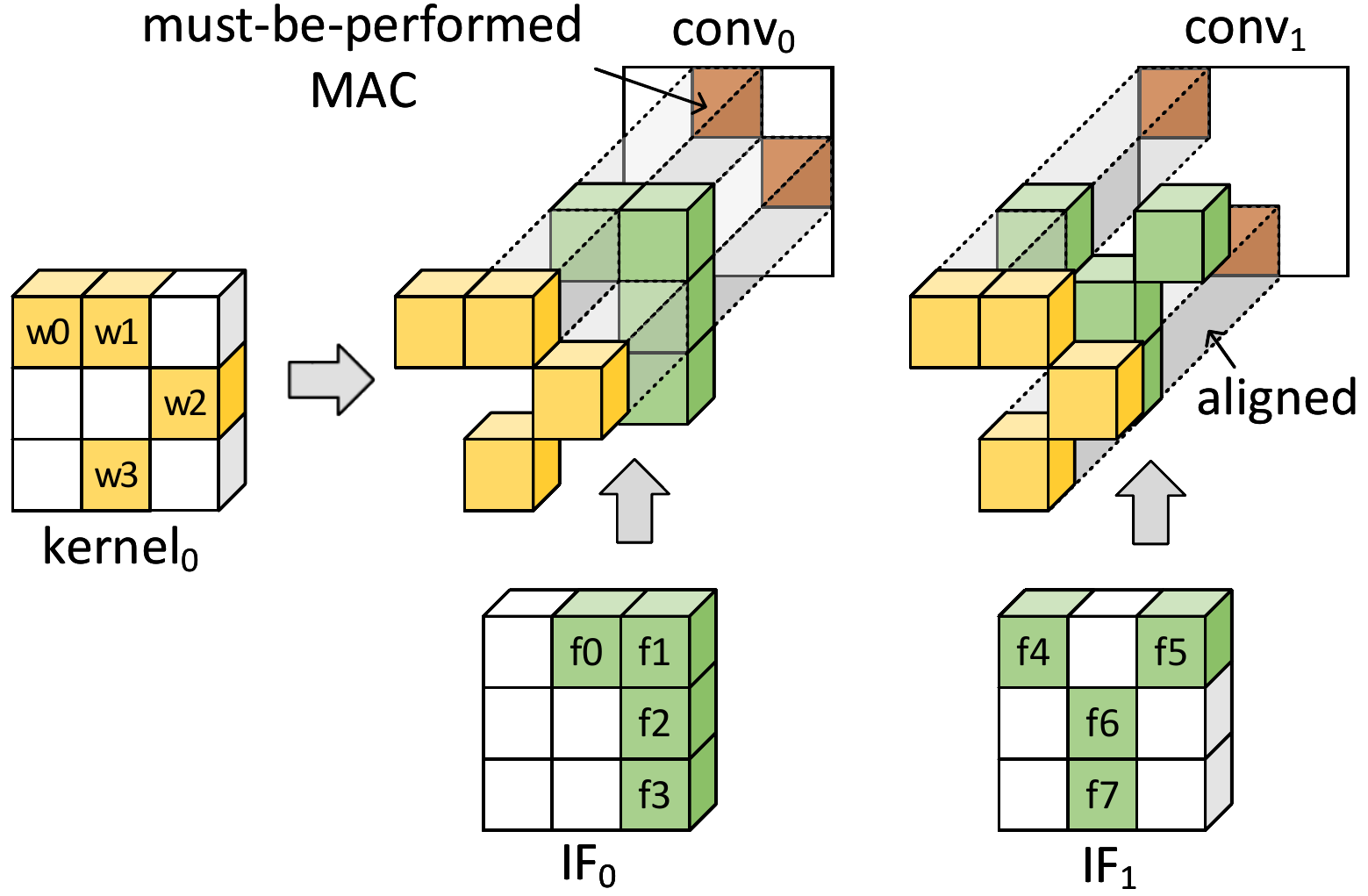}
    \caption{A demonstration of two convolutions with sparse feature and weight. The MAC operation must be performed only when the corresponding positions of the weight and feature are both non-zero, i.e. aligned-pair $(w, f)$. }
    \label{fig3}
\end{figure}

\begin{figure}[t]
    \centering
    \includegraphics[width=\columnwidth]{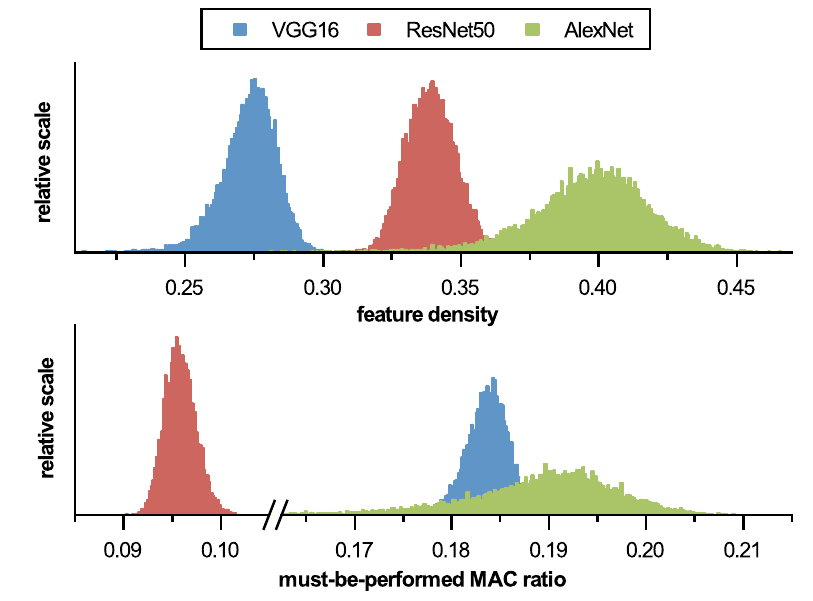}
    \caption{The distribution of feature density and must-be-performed MAC ratio on ImageNet dataset. The feature density is defined as the proportion of non-zero elements in all feature maps, and the must-be-performed MAC ratio is defined as the proportion of MAC operations with two non-zero operands (the aligned data pairs) in all convolutions.}
    \label{fig2}
\end{figure}

\subsection{Optimization with Sparsity}

\begin{table}[ht]
  \newcommand{\tabincell}[2]{\begin{tabular}{@{}#1@{}}#2\end{tabular}}
  \centering
  \begin{threeparttable}
  \caption{Weight and feature sparsity of different CNNs, represented by the percentage of zeros.}
  \label{tab2}
  \small
  \begin{tabular}{c|ccc}
    \specialrule{0.8pt}{0pt}{0pt}
              & AlexNet & VGG16 & ResNet50 \bigstrut \\ \cline{1-4}
    Average Weight Sparsity   & 64\%    & 68\%   & 76\% \bigstrut \\ 
    Average Feature Sparsity   & 61\%    & 72\%   & 66\% \bigstrut \\ 
    \specialrule{0.8pt}{0pt}{0pt}
  \end{tabular}
  \end{threeparttable}
\end{table}

Sparsity widely exists in modern deep CNNs, including weight sparsity and feature sparsity. 
For example, the intrinsic redundancy of deep CNNs allow us to prune majority of the weights with negligible accuracy loss by re-training the models \cite{19-han2015deep, 18-han2015learning}. 
Moreover, feature sparsity is introduced during inference by ReLU function that converts negative inputs to zeros.
The average weight sparsity of three typical CNNs are listed in~\cref{tab2}.
Because feature sparsity varies with different input images, we randomly select $50000$ images from ImageNet \cite{26-deng2009imagenet} and calculate their corresponding feature sparsity. 
Since all the MACs with zero operand(s) can be skipped without affecting the convolution result, only the MACs on aligned data pairs have to be computed and defined as must-be-performed MACs as shown in \cref{fig3}.
The distributions of feature density and must-be-performed MAC ratios of the three CNNs are depicted in \cref{fig2}.
According to these statistics, the sparsity of both weights and features is not trivial and provides a considerable opportunity to reduce computational cost and memory footprint.

Although the sparsity of CNNs is not trivial, its irregularity makes it difficult to effectively accelerate the executions of the sparse CNNs~\cite{15-20-zhang2016cambricon}.
\cref{fig3} demonstrates the fundamental challenge brought by such irregularity.
To eliminate all the unnecessary MAC operations (with one or two zero operands), only the aligned data pairs need to be sent to PEs.
As a result, although $conv_0$ and $conv_1$ share the same kernel, the actual weights they access in the kernel can be totally different due to the difference between the sparsity patterns of ${IF}_0$ and ${IF}_1$.
\cref{convnet} demonstrates that how such an irregularity brings challenge to systolic array.
The weights required by PE$_0$ ($w_{1,0}$) and PE$_1$ ($w_{0,0}$ and $w_{3,0}$) for convolution are different, which breaks the data transmission path inside the systolic array.
This characteristic also prevents the accelerators with explicitly planned dataflow \cite{11-jouppi2017datacenter,16-24-chen2017eyeriss,25-tu2017deep} from fully leveraging the network sparsity.
It also becomes challenging to optimize both data reuse and sparsity at the same time.

\begin{table}[ht]
  \newcommand{\tabincell}[2]{\begin{tabular}{@{}#1@{}}#2\end{tabular}}
  \centering
  \begin{threeparttable}
  \caption{Comparison of sparsity considerations on different CNN accelerators. Here $\cal{F}$ means only feature sparsity is considered and $\cal{W}$ means only weight sparsity is considered.}
  \label{tab3}
  \small
  \begin{tabular}{m{2.4cm}<{\centering}|m{1.8cm}<{\centering}|m{1cm}<{\centering}|m{1.8cm}<{\centering}}
    \specialrule{0.8pt}{0pt}{0pt}
    Accelerator & Gate MAC & Skip MAC & \tabincell{c}{Skip Buffer \\ or DRAM \\ Accesses} \bigstrut \\
    \hline
     TPU \cite{11-jouppi2017datacenter}     & Unmentioned & - & Unmentioned   \bigstrut   \\
    \hline
    Eyeriss \cite{16-24-chen2017eyeriss}   & $\cal{F}$ & - & $\cal{F}$ \bigstrut  \\ 
    \hline
    Cnvlutin \cite{21-albericio2016cnvlutin}  & $\cal{F}$  & $\cal{F}$ & $\cal{F}$ \bigstrut  \\ 
    \hline
    Cambricon-X \cite{15-20-zhang2016cambricon} & $\cal{W}$  & $\cal{W}$ & $\cal{W}$ \bigstrut  \\
    \hline
    \cite{kung2018packing}  & $\cal{W}$  & $\cal{W}$  & $\cal{W}$  \bigstrut  \\ 
    \hline
    Cambricon-S \cite{22-zidong2018cambricon-s} & $\cal{F}+\cal{W}$ & $\cal{F}+\cal{W}$ & $\cal{F}+\cal{W}$ \bigstrut  \\ 
    \hline
    EIE \cite{Han2016EIE}           & $\cal{F}+\cal{W}$ & $\cal{F}+\cal{W}$ & $\cal{F}+\cal{W}$ \bigstrut  \\ 
    \hline
    SCNN \cite{17-parashar2017scnn} & $\cal{F}+\cal{W}$ & $\cal{F}+\cal{W}$ & $\cal{F}+\cal{W}$ \bigstrut   \\ 
    \hline
    SparTen \cite{gondimalla2019sparten} & $\cal{F}+\cal{W}$ & $\cal{F}+\cal{W}$ & $\cal{F}+\cal{W}$ \bigstrut   \\ 
    \hline
    \SSQ Engine                     & $\cal{F}+\cal{W}$ & $\cal{F}+\cal{W}$ & $\cal{F}+\cal{W}$ \bigstrut  \\
    \specialrule{0.8pt}{0pt}{0pt}
  \end{tabular}
  \end{threeparttable}
\end{table}

\Cref{tab3} summarizes the sparsity strategies explored by existing accelerators.
For example, TPU does not explore any of the two sparsity strategies in systolic arrays \cite{11-jouppi2017datacenter}, since each zero would inevitably occupy a PE.
Eyeriss only exploits the feature sparsity \cite{16-24-chen2017eyeriss}.
By utilizing feature sparsity, Cnvlutin achieves performance improvement by skipping the operations with zero elements in feature maps \cite{21-albericio2016cnvlutin}.
Cambricon-X \cite{15-20-zhang2016cambricon} and \cite{kung2018packing}, however, only exploit weight sparsity within the network trained by their proposed pruning algorithm.
Cambricon-S fully deploys the sparsity of both features and weights, but only supports them at coarse granularity and requires additional pruning algorithms \cite{22-zidong2018cambricon-s}.
EIE also exploits these two types of sparsity \cite{Han2016EIE} but is only designed for fully-connected networks.
SCNN exploits these two types of sparsity on both convolutional and fully-connected layers \cite{17-parashar2017scnn}. However, SCNN requires lots of coordinate computations and introduces significant overhead.
As an evidence, SCNN only achieves $79\%$ of the speed but consumes $33\%$ more energy when processing dense CNNs \cite{17-parashar2017scnn} .
SparTen \cite{gondimalla2019sparten} supports sparse vector-vector multiplication and load balance optimization to improve the hardware utilization, but the required additional computation resources degrade the energy efficiency significantly.

\section{S\raise0.7ex\hbox{\normalsize 2}\ Engine Architecture}\label{sec4}

\subsection{Architecture Overview}

\begin{figure}[t]
    \centering
    \includegraphics[width=0.9\columnwidth]{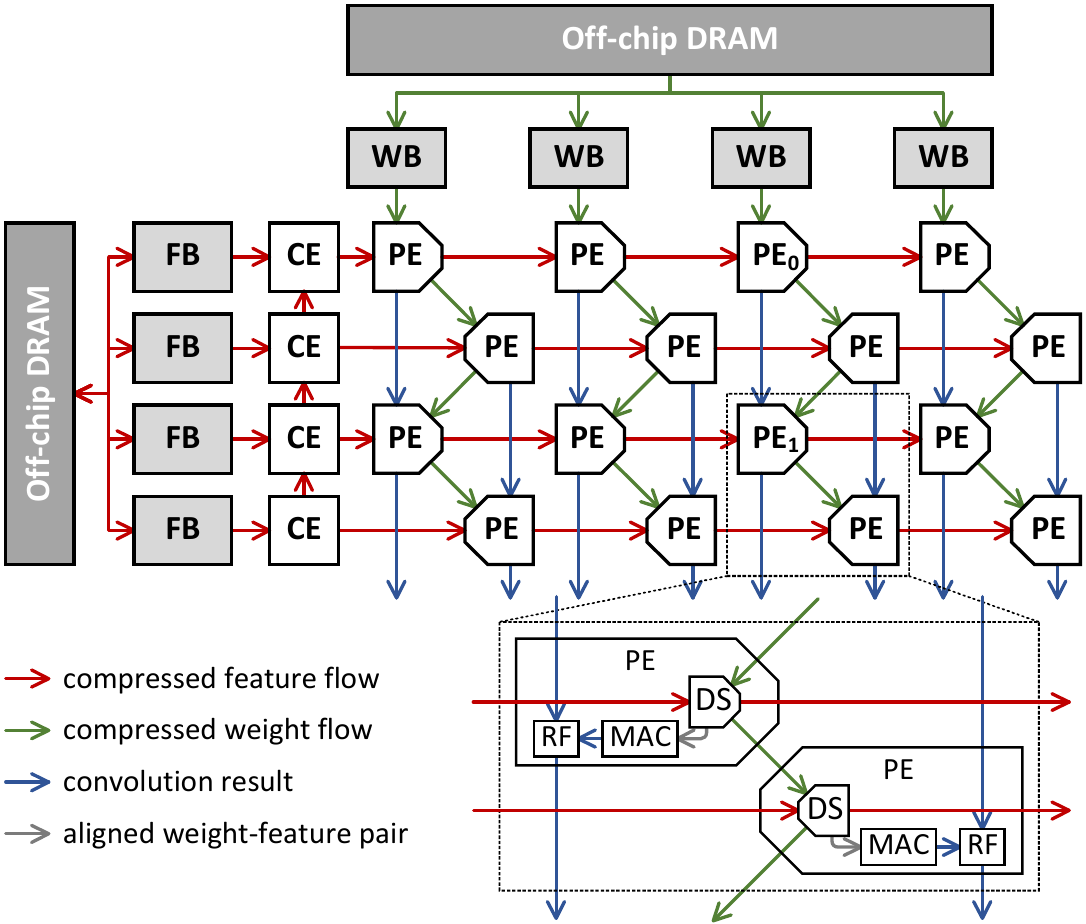}
    \caption{Architecture overview of \SSQ Engine and demonstration of internal dataflow.}
    \label{arch}
\end{figure}

The top-level architecture of \SSQ Engine and the schematic of its internal dataflow are shown in \cref{arch}. \SSQ Engine consists of a homogeneous PE array and an associated CE (collective element) array. Sparse features are compressed and stored in feature buffer (FB) while sparse weights are compressed and stored in weight buffer (WB). 

As aforementioned, weight stationary dataflow adopted in TPU \cite{11-jouppi2017datacenter} prevents it from deploying the sparsity. Therefore, \SSQ Engine utilizes an output stationary dataflow as shown in \cref{arch}.
Similar to \cref{convnet}, each PE undertakes the computation of a separate convolution while the compressed feature and weight flow among the systolic array in two directions in order to realize data reuse.
The sparse features are fetched by CE array and sent to PE array. The feature overlap is processed between different PE rows to achieve overlap reuse.
The data transmission path shown in \cref{arch} allows the \SSQ Engine to transmit the convolution results out of the systolic array faster compared to the na\"ive design adopted in \cite{10-wei2017automated}.

In this work, we enhance the PE design from the na\"ive design \cite{10-wei2017automated, 11-jouppi2017datacenter} to exploit the sparsity of both weight and feature. 
Our PE can dynamically select the aligned data pairs from the two dataflows that passing across it for sparse convolution. 
As shown in \cref{arch}, each PE can be decomposed into three components: \underline{D}ynamic \underline{S}election (DS), \underline{M}ultiplication and \underline{AC}cumulation (MAC), \underline{R}esult \underline{F}orwarding (RF).
DS component performs dynamic selection and then sends aligned feature-weight pair to MACs. 
The design of MAC component is trivial as it just simply achieves the multiplication and accumulation.
The function of RF component is illustrated with PE$_0$ and PE$_1$ marked in \cref{arch}.
Due to the irregularity of sparsity, the workload allocated to each PE might not be equal. For example, PE$_1$ might generate the convolution results before PE$_0$.
To guarantee the convolution results are transmitted out from the systolic array sequentially, in addition to the na\"ive design, the RF component in PE$_1$ needs to stall and wait until the convolution results from PE$_1$ have been forwarded.

Similar to the na\"ive design shown in \cref{convnet}, the convolution process would be projected to \texttt{GEMM} (general purpose matrix multiplication) operation for the PE array to process.
However, different from the na\"ive \texttt{im2col()} operation provided in Caffe \cite{jia2014caffe}, the three dimensional input feature map is divided into groups and then reshaped into one-dimensional vector at this granularity.
Such a reshaping manner is adopted for the convenience of deploying overlap reuse which would be detailed illustrated in \cref{sec4_4}.
After being reshaped into one-dimensional data (illustrated in \cref{convnet}), the sparse data are further compressed before feeding into PE array as demonstrated in \cref{arch}.

\begin{figure}[t]
    \centering
    \includegraphics[width=\columnwidth]{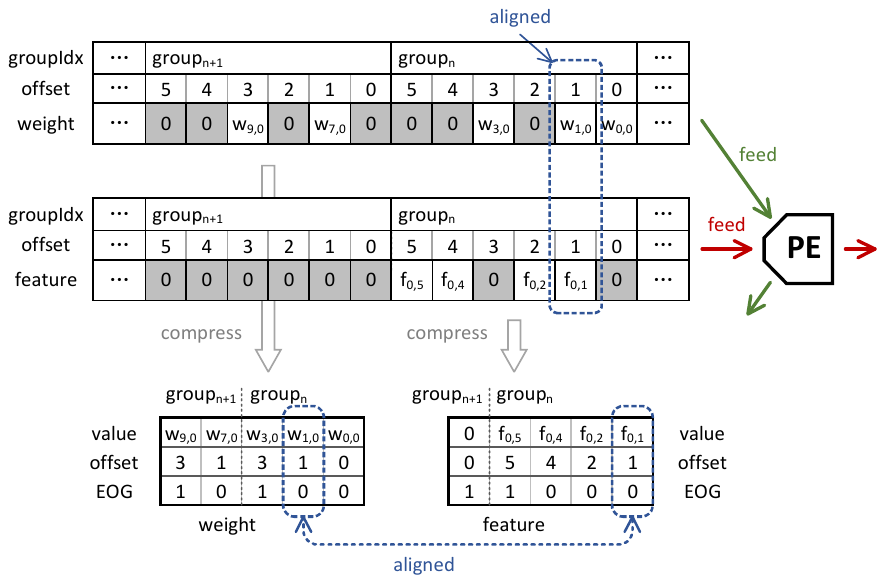}
    \caption{A toy example to illustrate the dataflow compression.}
    \label{compress}
\end{figure}

\subsection{Dataflow Compression}\label{sec4_2}

To select all the aligned data pairs, e.g. $(f_{0,1}, w_{1,0})$ in \cref{convnet}, from the compressed feature and weight flow, the indices of non-zero feature and weight need to be extracted and compared with each other in PEs.
However, several popular sparse formats are not suitable for this purpose because they usually introduce additional overheads.
COO \cite{Vuduc2003COO} format encodes two absolute indexes for each element, which contains redundant information and cannot represent an arbitrary large matrix with limited bit width for offset.
The CSC format \cite{Han2016EIE} stores the number of zeros between two non-zeros, therefore non-zeros accessing has to perform coordinate transformation from the compressed format.
In this work, as a variation of COO, Enhanced COO format (ECOO) is introduced to overcome these limitations.
Instead of achieving a higher compression ratio, the primary goal of adopting ECOO format is to simplify the design of DS.

As described before, since the input data is reshaped at the granularity of groups, the one-dimensional vectors fed into the systolic array (feature or weight) have been naturally divided into groups as shown in \cref{compress}, where the group length is fixed at $6$ in this example.
The absolute position of each element inside each group is stored as \texttt{offset}. 
To avoid the mismatch of feature and weight elements between different groups, an extra \texttt{EOG} (end-of-group) flag is defined at the last element of each group. 
If all elements in a group are zeros, one zero is kept and marked as \texttt{EOG} as a placeholder. 
Hence, ECOO format could be denoted as triplets (\texttt{value}, \texttt{offset}, \texttt{EOG}).
As we have observed that $4$ bits are enough to represent \texttt{offset} which is also used in several previous works \cite{19-han2015deep,Han2016EIE}, the group length is set to $16$ in this work. 
Additional $1$ bit is required for \texttt{EOG} and each nonzero feature would be represented by 13 bits in total. Another 1 bit is required to encode nonzero weight ($14$ bits in total) to represent the end-of-kernel. 
As shown from \cref{compress}, the aligned weight-feature pair, e.g. $(w_{1,0},f_{0,1})$, would have the same \texttt{offset} after being compressed. 
In Section \ref{sec4_3}, we will show that this property makes it easier for the DS component to select all the aligned weight-feature pairs from the compressed dataflow.

\begin{figure}[t]
    \centering
    \includegraphics[scale=0.9]{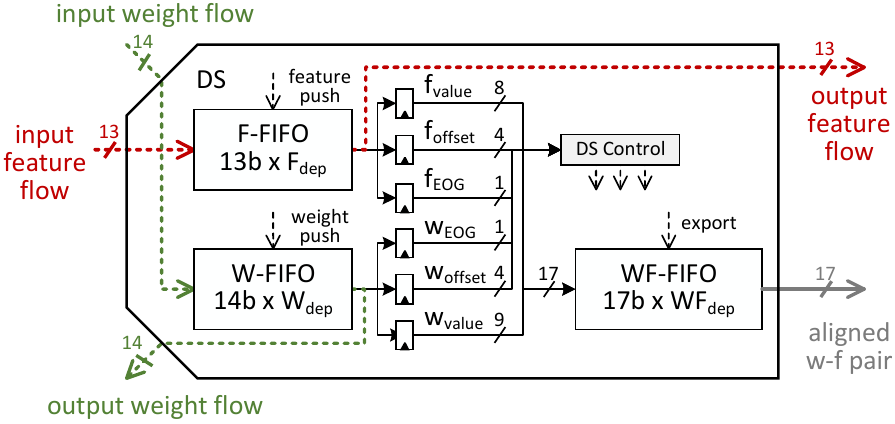}
    \caption{Datapath and functional components of DS design, where \texttt{w-f} represents weight-feature.}
    \label{DSarch}
\end{figure}

\subsection{Dynamic Selection Component} \label{sec4_3}
By adopting the ECOO format described above, DS component only needs to select all weight-feature pairs with the same offset and feed them into MACs.
The selecting process is scheduled by the controller to find the weight-feature pairs with the same offset.
As illustrated in \cref{DSarch}, weight and feature flows are first buffered in W-FIFO and F-FIFO, respectively.
Then the controller could select the aligned weight-feature pairs (stored in $w_{\text{value}}$ and $f_{\text{value}}$ respectively) according to their \texttt{offset} (stored in $w_{\text{offset}}$ and $f_{\text{offset}}$ respectively) and \texttt{EOG} (stored in $w_{\text{EOG}}$ and $f_{\text{EOG}}$ respectively), and store them into WF-FIFO.
After that, the weight and feature flows are exported to the adjacent PEs for data reuse among the entire systolic array, as demonstrated in \cref{arch}.

\begin{figure}[t]
    \centering
    \includegraphics[width=\columnwidth]{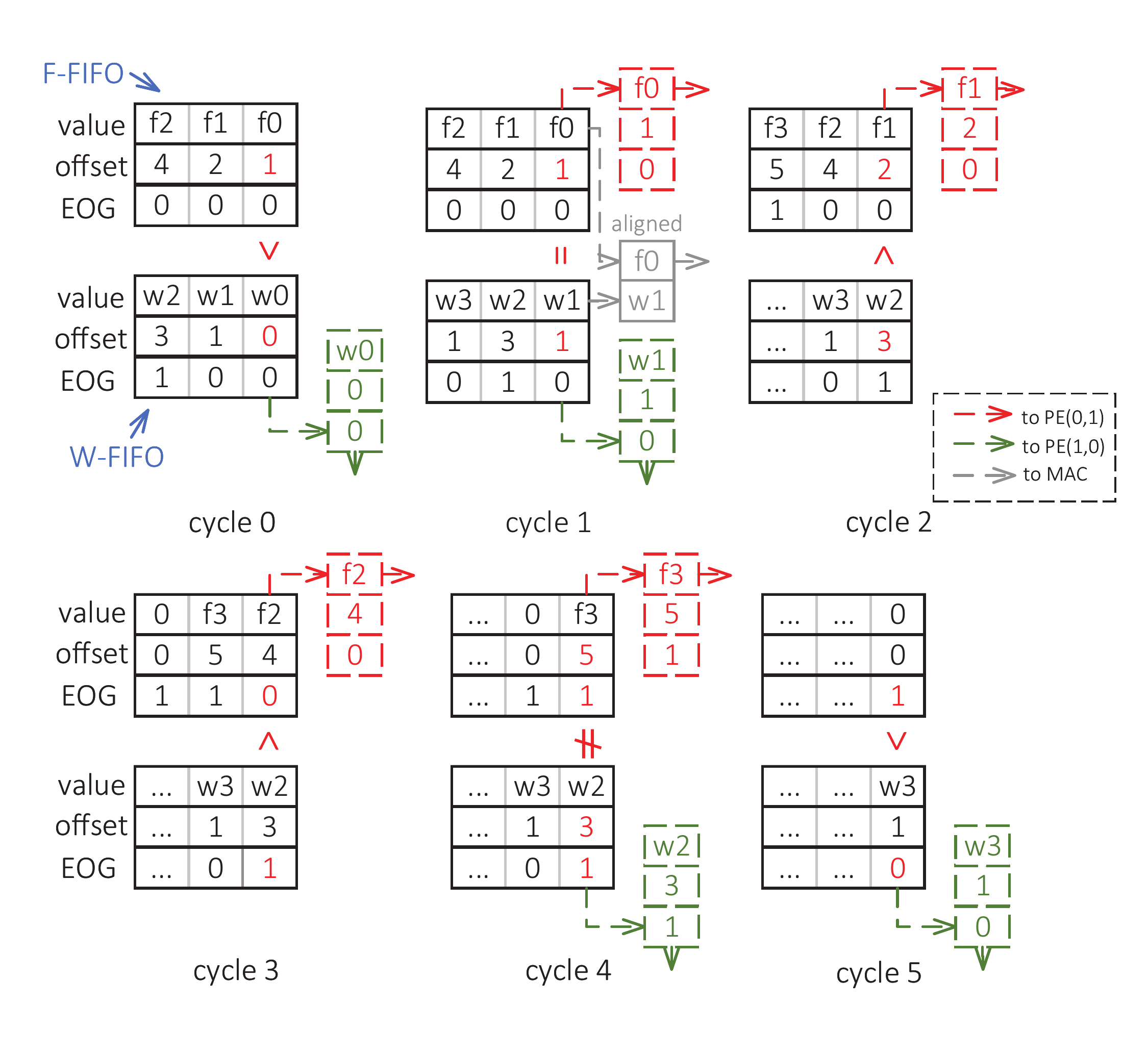}
    \caption{ Demonstration of the dynamical selection process (from cycle\textsubscript{0} to cycle\textsubscript{5}) using a toy example in PE(0,0).}
    \label{DSprocess}
\end{figure}

A detailed illustration of this dynamical selection process with the toy example in \cref{compress} is shown in \cref{DSprocess}. 
In this example, utilizing the ECOO format can significantly simplify the selection process.
For the convenience of the following description, we define the \texttt{push} of dataflow as an action that transmitting one data from FIFO to both the registers and the succeed PE in its transmission path. For example, weight flow is pushed in \texttt{cycle}$_0$ while feature flow is pushed in \texttt{cycle}$_2$.
In \texttt{cycle}$_0$, feature flow is pushed since w$_{\text{offset}}$ $<$ f$_{\text{offset}}$.
Then in \texttt{cycle}$_1$, because of w$_{\text{offset}}$ $=$ f$_{\text{offset}}$, the weight-feature pair stored in registers is aligned and is sent to MAC. Meanwhile, both weight and feature flow are pushed to process the subsequent data.
In \texttt{cycle}$_3$, weight meets \texttt{EOG} but feature does not. Therefore, feature flow is pushed until its meets \texttt{EOG} too in \texttt{cycle}$_4$. After that, DS begins to process the next group of data in \texttt{cycle}$_5$.

This toy example also illustrates how \SSQ Engine achieves speedup by deploying network sparsity.
As demonstrated in \cref{DSprocess}, the processing of one group (\texttt{group}$_n$) is completed in five cycles while the two groups (\texttt{group}$_n$ and \texttt{group}$_{n+1}$) is processed in seven cycles.
It can be seen that more cycles (six cycles for each group) would be required to process them in a na\"ive design \cite{11-jouppi2017datacenter,10-wei2017automated}.
Moreover, because there is only one aligned weight-feature pair is selected in this process, the MAC component would be idle for most of time if it runs at the same frequency as DS component.
It can be further inferred that a higher utilization of MAC component and a higher throughput can be achieved when DS runs at a higher frequency than MAC component.
The impact of the frequency ratio between these two components on throughput is thoroughly evaluated in \cref{sec6}.

On the other hand, as demonstrated in \cref{DSprocess}, such a dynamical selection process can also result in noncontinuous movement of both feature and weight flows.
The performance of the succeed PE on data transmission path will be degraded.
Hence, the FIFOs inserted in DS component can provide such a discontinuity in order to make the whole systolic array run smoothly.
Based on our observations, several tens of bits are enough to implement the required FIFOs (represented by registers). 
The impact of FIFO size on total performance will be evaluated in \cref{sec6}.

\subsection{Collective Element}\label{sec4_4}
\begin{figure}[t]
    \centering
    \includegraphics[width=\columnwidth]{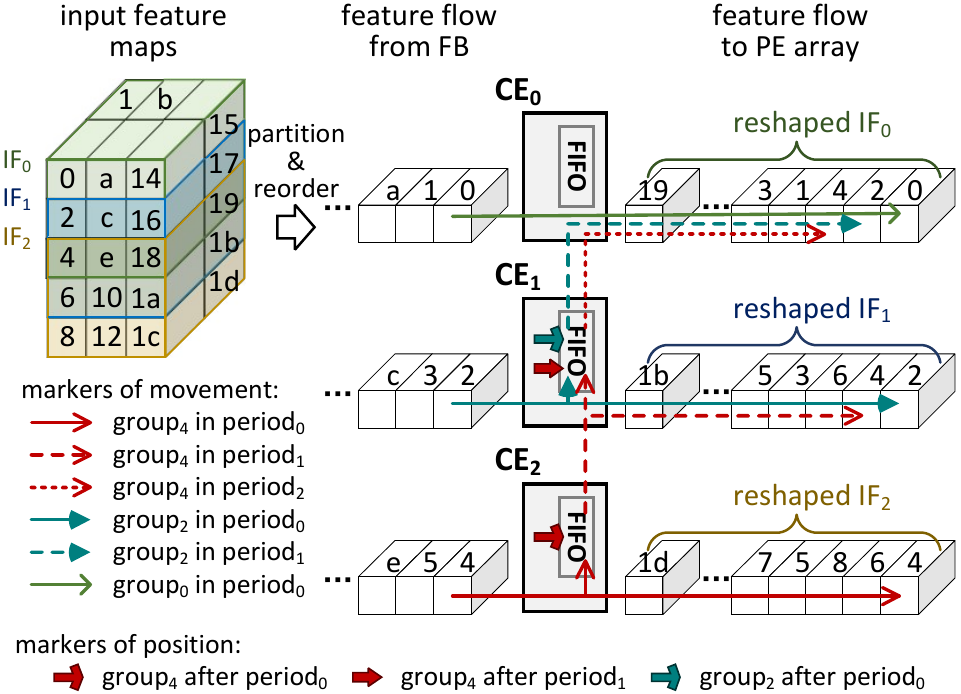}
    \caption{Internal data transmission of CE array. }
    \label{CEdataflow}
\end{figure}

CE array is designed to exploit the overlap reuse in adjacent rows of PE array as shown in \cref{convnet}.
The compressed data groups are fed into CE array and broadcasted according to the internal data transmission path as shown in \cref{CEdataflow}.
The input three-dimensional data (both feature and weight) are divided into groups along the channels (cubes in Fig.8), and each group contains up to 16 elements. After that, they are reshaped as one-dimensional dataflow and fed into different rows of PE array.
    
Considering a convolutional layer with a kernel size of $3 \times 3$ and a stride of $1$, as shown in \cref{CEdataflow}, the feature maps required for the three convolutions, \texttt{IF}$_0$, \texttt{IF}$_1$ and \texttt{IF}$_2$ is overlapped with each other. 
When this layer is processed on a na\"ive design shown in \cref{convnet}, the overlapped part of these three feature maps would be stored in three separate FBs as three copies, resulting in a waste of memory storage.
On the other hand, since the same group of data would be loaded from the three FBs separately, unnecessary power overheads are introduced by these repeated buffer accesses.

The working mechanism of CE array is illustrated in \cref{CEdataflow}.
The data transmission procedure is divided into three periods, where data movement path and data groups are highlighted.
In \texttt{period}$_0$, \texttt{CE}$_2$ loads \texttt{group}$_4$ from FB and then send this group to PE array while holding a copy of the group in its internal FIFO.
During the same period, \texttt{group}$_2$ is sent to PE array by \texttt{CE}$_1$ in the same manner.
In the next period, \texttt{group}$_4$ is loaded from the internal FIFO in \texttt{CE}$_2$ and sent to PE array by \texttt{CE}$_1$.
\cref{CEdataflow} shows that each CE only holds one group of data at a time.
Since the input feature map are divided into groups along channels, the above dataflow generation manner could guarantee that such a CE array could process all scales of CNNs by using internal FIFOs with fixed depth.

In such an approach, CE array mainly involves lots of internal FIFO accessing (small register files) instead of frequently FB accessing (large SRAM) so that its energy efficiency could be improved accordingly.
Meanwhile, each CE only stores one group of data as shown in \cref{CEdataflow}, i.e. \texttt{CE}$_2$ only stores \texttt{group}$_4$ in its internal FIFO, so that the deployed CE array could make \SSQ Engine very memory efficient.

The CE array runs at the same frequency as DS component and the evaluation in \cref{sec6} shows that it does not cause a performance bottleneck of \SSQ Engine.
Different from the strategy adopted in \cite{xin2017cosy} which introduces auxiliary structure in each PE, the CE array adopted in this design can deploy the overlap reuse with negligible extra resources. 
Without loss of generality, our CE array can be also applied to na\"ive systolic array shown in \cref{convnet}.

\subsection{Mixed-precision Data Processing}

\begin{figure}[t]
    \centering
    \includegraphics[scale=1]{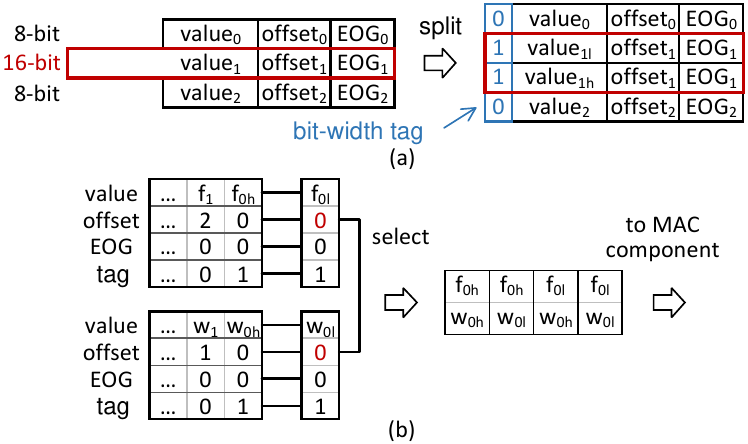}
    \caption{(a) the unified representation for both $8$-bit and $16$-bit data with an extra flag bit. (b) disassembling $16$-bit multiplication into four $8$-bit multiplications.}
    \label{fig10}
    \vspace{-3pt}
\end{figure}

\SSQ Engine can further support mixed-precision computing at a fine-grained manner. 
The work in \cite{14-park2018energy} introduces extra data path to process higher precision data. However, our PE is only designed with an 8-bit data path.
During the dataflow compression procedure, data will be divided into two regions, $8$-bit and $16$-bit ones, according to the given threshold. Then a $8$-bit data will be marked with tag $0$, and a $16$-bit data \texttt{value}$_1$ has to be split into two $8$-bit data and marked with tag $1$, as shown in \cref{fig10}(a).
After that, the two $8$-bit data could be fed into PEs for normal processing.
\cref{fig10} (b) also demonstrates a special situation when two $16$-bit data meet at the same PE. This situation could be solved by dividing the data into four pairs and feeding the pairs into the PEs.
It is obvious that the mixed-precision data processing in our approach will not degrade the throughput of PE array, as validated by the experimental results in \cref{sec6}.

\section{Experimental Methodology}\label{sec5}

The kernel parts of \SSQ Engine, including PE, CE and FIFO, are implemented as Verilog RTL and synthesized by Synopsys DC with Global Foundry $14$nm LP FinFET technology.
Gate level simulations are performed by Synopsys VCS simulator by setting the frequency of MAC component as $500$MHz. 
The area cost and energy consumption of PE, CE and FIFO are analyzed by PrimeTime.
The area and energy consumption of buffers (both WB and FB) are estimated by PCACTI \cite{shafaei2014fincacti} while the energy consumption of off-chip DRAM is estimated by CACTI \cite{muralimanohar2009cacti}.

\subsection{Performance Evaluation}

The utilized sparse CNN models are trained by the pruning algorithm proposed in \cite{18-han2015learning}.
An in-house compiler is designed with C++ language to translate the sparse CNN models into compressed dataflow for \SSQ Engine simulations.
Meanwhile, the required buffer capacity and the total amount of buffer accessing is evaluated by analyzing the generated compressed dataflow.
A cycle-by-cycle accurate simulator is implemented with C++ language, which is practical to evaluate the performance of \SSQ Engine under various configurations.
Our simulator models the cycle-by-cycle behaviors of each atomic component, including (1) RF/FIFO in PE and CE, (2) MAC in PE, (3) DS in PE, (4) WB and FB, etc.
The generated dataflow is fed into the simulator to obtain the involved execution cycles as well as the statistics on the behaviors of the atomic components for latency and energy efficiency estimation.

\subsection{Architecture Configurations}

Several kinds of configurable parameters are explored to comprehensively evaluate the performance of \SSQ Engine.
First, inheriting the modularity and expandability of systolic architecture, \SSQ Engine can be configured as different scales for various throughput requirements.
Moreover, both the size of FIFOs and frequency ratio of DS-MAC component in each PE can be also configured as mentioned in \cref{sec4}. 
The na\"ive systolic array with output stationary dataflow demonstrated in \cref{convnet} is adopted as a baseline, which can be also basically regarded as the performance of TPU.
Similar to SCNN~\cite{17-parashar2017scnn}, the na\"ive systolic array is configured with total $2$MB SRAM for FB and WB. 
This capacity is sufficient to hold $66$ out of $71$ convolution layers we evaluated. 
Since both feature and weight are compressed while the required buffer capacity can be further reduced by CE array, $1$MB SRAM is sufficient for \SSQ Engine to hold $68$ out of $71$ layers.
With a similar conclusion in \cite{17-parashar2017scnn}, the DRAM bandwidth is configured as $50$GB/s which will not become as a performance bottleneck.
Aiming to estimate the speedups brought by sparsity, the na\"ive systolic array runs at the same frequency as the MAC component in \SSQ Engine. 
Besides, the na\"ive design adopts the same convolution mapping strategy as \SSQ Engine, which provides a fair comparison.
In order to evaluate the performance degradation incurred by the blocking behaviors of finite-depth FIFO in PE, an infinite-depth FIFO is adopted as an ideal situation which is corresponding to the upper bound of performance.

\subsection{Benchmarks}

\SSQ Engine is evaluated on both synthetic and actual CNN models.
Synthetic models are utilized to analyze some typical characteristics of \SSQ Engine while actual models are adopted to evaluate the realistic impacts of different design aspects.
A series of CNN models are synthesized by different designated sparsity levels both on features and weights to evaluate the sensitivity of our design as the sparsity changes.
Similar to \cite{18-han2015learning}, actual sparse CNN models are generated by training on ImageNet\upcite{26-deng2009imagenet} and pruning with a neural network distiller \cite{distiller}.
Since the feature maps are resulted from input images, their sparsity can vary significantly with different input images and greatly affect the evaluation results.
The ImageNet is divided into three subsets according to their resulted different feature sparsity (maximum, average and minimum) for comprehensive evaluation.
If not mentioned, the following evaluation results are obtained by average feature sparsity.

\begin{figure}[t]
    \centering
    \includegraphics[scale=0.65]{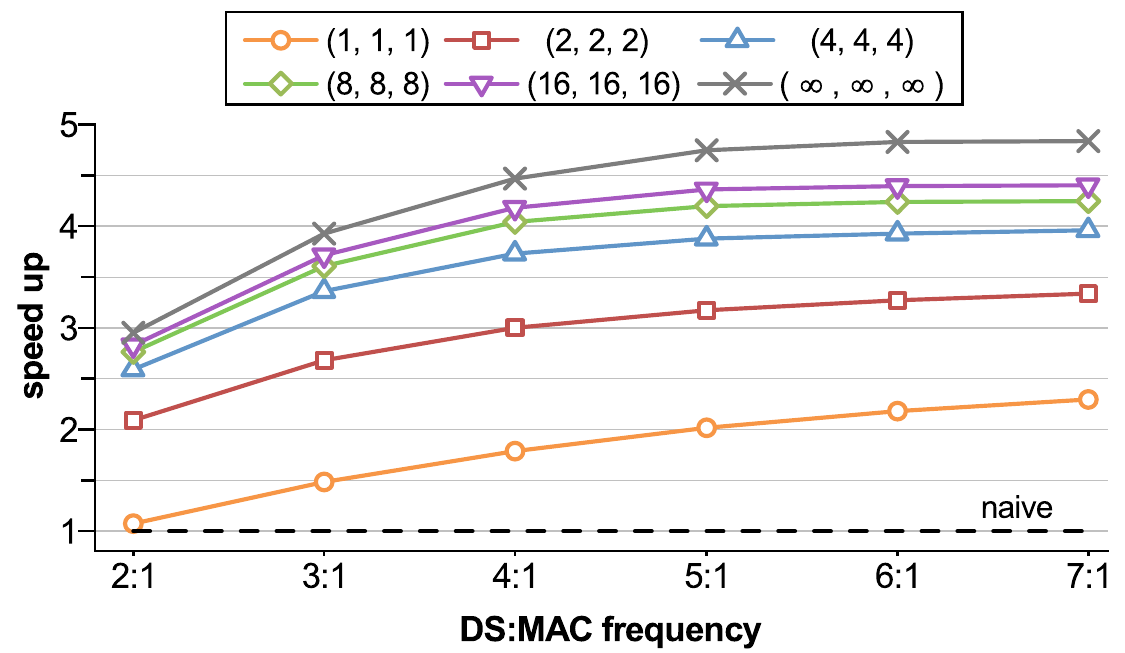}
    \caption{Speedups with different FIFOs depth and DS:MAC frequency ratio.}
    \label{fig11}
\end{figure}

\section{Evaluation Results}\label{sec6}

This section first evaluates the characteristics of PE array in \SSQ Engine separately, including speedup under different configurations, and sensitivity to different sparsity levels. 
Then, the benefits brought by adopted CE array in \SSQ Engine are separately measured on actual models.
At last, the complete \SSQ Engine is evaluated on different actual sparse CNN models to demonstrate the performance improvement in speedup, energy and area efficiency versus the na\"ive systolic array.

\subsection{Design Space Exploration}\label{sec6_1}

As discussed before, both FIFOs depth and frequency of DS component can largely affect the throughput of \SSQ Engine.
These impacts are evaluated on actual sparse models by fixing the scale of PE array as $16\times16$.
The evaluation results also provide guidance for parameters selection in the following experiments.

\cref{fig11} illustrates the average speedups of PE array when evaluating AlexNet/VGG16/ResNet50 under different configurations: FIFOs depth and frequency ratio of DS:MAC.
Several typical combinations of FIFOs depth are selected and evaluated while the performance of other settings can be inferred from these results.
Corresponding to the symbols in \cref{DSarch}, the FIFOs depth are denoted as (\texttt{W}$_\text{dep}$, \texttt{F}$_\text{dep}$, \texttt{WF}$_\text{dep}$) in \cref{fig11}.
Also, the infinite depth of FIFOs are denoted as $(\infty,\infty,\infty)$.

The results in \cref{fig11} show that the benefits brought by higher DS frequency and deeper FIFOs diminish on marginal.
When FIFOs depth is increased from (2,2,2) to (4,4,4), there are about $1.2\times$ speedups achieved.
When FIFOs depth is further increased to (8,8,8), only about $1.1\times$ additional speedups are obtained.
As a result, relatively small FIFOs with affordable area overhead are sufficient for data buffering. 
The similar situation could be found when varying DS:MAC frequency ratio.
There are about $1.5\times$ speedups achieved when DS:MAC frequency ratio is doubled from $2$:$1$ to $4$:$1$.
If the ratio is further doubled to 8:1, only about $1.1\times$ additional speedups could be obtained.
Since the speedups are almost saturated when frequency ratio reaches to $4$:$1$, i.e. DS component runs at $2000$MHz, DS:MAC frequency ratio is set as $4$:$1$ in the following experiments. Meanwhile, such a frequency ratio can be certainly realized by frequency division with a single clock domain.

\begin{figure*}[t]
    \centering
    \includegraphics[scale=1.4]{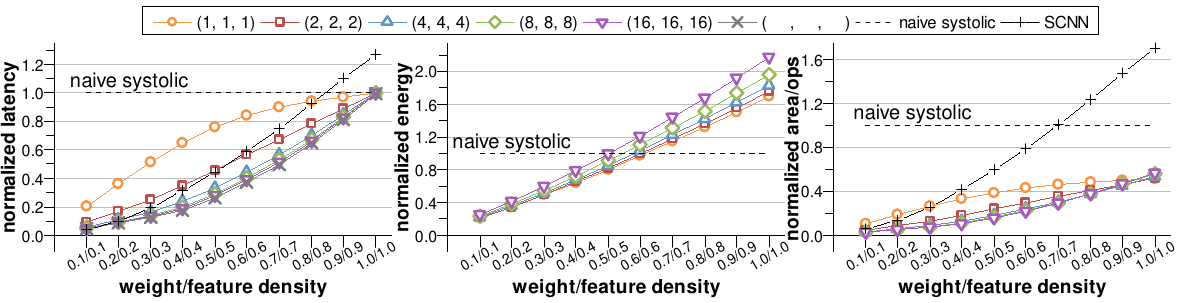}
    \caption{Normalized latency, energy and area efficiency with different sparsity levels (\texttt{density}) and different FIFOs depth.}
    \label{fig12}
\end{figure*}

\subsection{Sensitivity to CNN Sparsity}\label{sec6_2}

Since sparsity levels usually affect the performance of CNN accelerators, we also evaluate different sparsity levels in actual CNN models. Aiming to compare with SCNN, our PE array is fixed as $32 \times 32$.
As shown in \cref{fig12}, a series of synthetic AlexNet models are evaluated by varying the sparsity levels both on features and weights from $10\%$ to $100\%$. The sparsity level is defined as \texttt{density}, i.e. the percentage of non-zeros in features or weights.
In general, our approach could obtain significant speedups compared with the na\"ive design because of the benefits from sparsity.
Additionally, we could achieve better performance or efficiency than SCNN in most scenarios.

On-chip energy is evaluated separately under different configurations as shown in \cref{fig12}. Our proposed PE array achieves better on-chip energy efficiency versus the na\"ive design when the density is lower than $0.5$/$0.5$.
Additionally, the energy efficiency would be further improved by introducing CE array which will be illustrated in \cref{sec6_5}.
For area efficiency, a metric of \texttt{area/ops}, i.e. the required chip area per operation, is defined to evaluate the area overhead introduced by DS component.
Considering that smaller SRAM is required by \SSQ Engine, its area consumption will be even smaller than the na\"ive design (see the area breakdown in \cref{tabCompare}) which leads to a significant improvement in area efficiency as shown in \cref{fig12}.
Since a large portion of resources are required by the involved crossbar and accumulator buffers in SCNN, our proposed PE array could have much more benefits on area efficiency than SCNN.

In conclusion, our adopted PE array behaves better than SCNN and the na\"ive design under most of the configurations when feature/weight density is lower than $0.5/0.5$. Such a density can be easily satisfied by actual CNNs according to the statistics in \cref{tab2} and \cref{fig2}.
However, the utilized non-zero patterns of features and weights are uniformly distributed, which is different from the actual situation where the large data tends to concentrate \cite{22-zidong2018cambricon-s}.
In the following experiments, we will evaluate \SSQ Engine by using actual sparse CNN models.

\subsection{Sensitivity to 16-bit Data Ratio}

\begin{figure}[t]
 \centering
 \centerline{\includegraphics[scale=0.7]{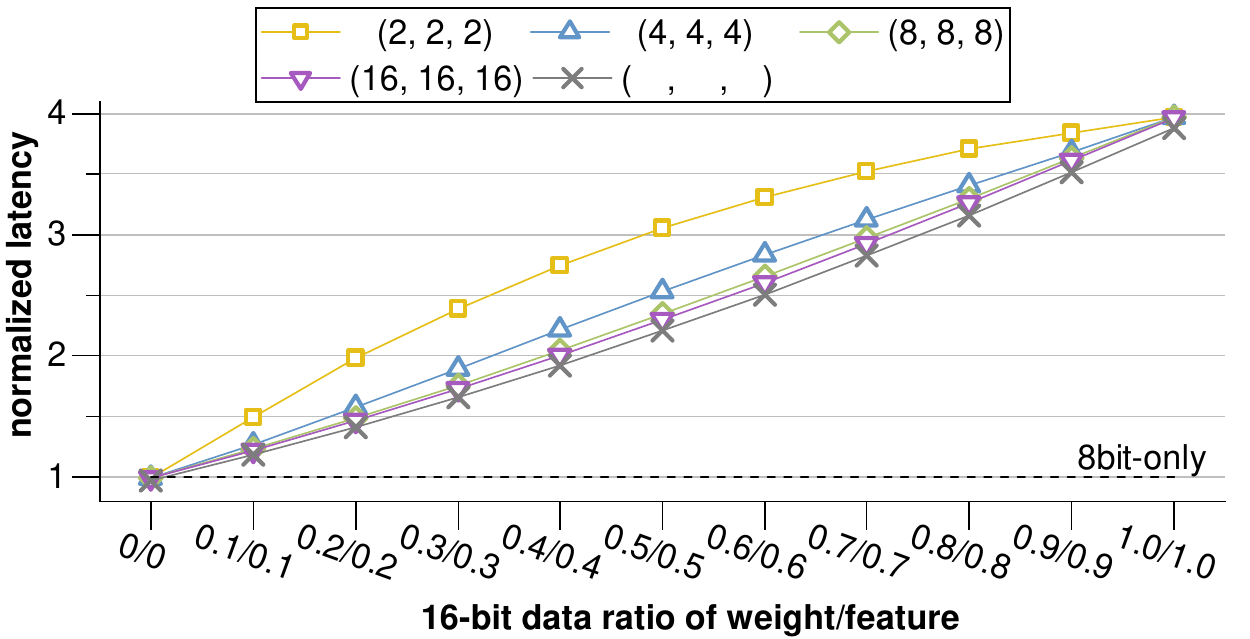}}
 \caption{Normalized latency versus $16$-bit data ratio for different FIFOs depth.}
 \label{fig13}
\end{figure}

Similar to the sensitivity to sparsity, the proposed mixed-precision data processing strategy is evaluated on a series of generated dense AlexNet models with $16$-bit data ratio growing from $10\%$ to $100\%$ where the remaining data are $8$-bit.
Since $8$-bit and $16$-bit data are processed in the same datapath, the results in \cref{fig13} show that our strategy is robust to various $16$-bit data ratio.
The evaluation in \cref{tab4} shows that our strategy can process mixed precision data more efficiently than the work in \cite{14-park2018energy}.

\begin{table}
  \centering
  \caption{Required additional running cycles of mixed-precision data processing compared with $8$-bit only strategy for different FIFOs depth, which are evaluated on the generated model.}
  \label{tab4}
  \small
  \begin{tabular}{|m{0.11\columnwidth}<{\centering}|*{3}{m{0.11\columnwidth}<{\centering}|}m{0.14\columnwidth}<{\centering}|c|}
    \hline
    $16$-bit Ratio      & (2,2,2) & (4,4,4) & (8,8,8) & (16,16,16) & \cite{14-park2018energy}  \bigstrut \\ \hline
    $3.5\%$             & $16.3\%$   & $9.1\%$    & $8.4\%$    & $8.2\%$       & $10\%$  \bigstrut \\ \hline
    $5\%$               & $24.1\%$   & $13.1\%$   & $11.9\%$   & $11.7\%$      & $\sim 20\%$ \bigstrut \\ \hline
  \end{tabular}
\end{table}

\subsection{Memory Efficiency}

\begin{figure}[t]
    \centering
    \includegraphics[width=\columnwidth]{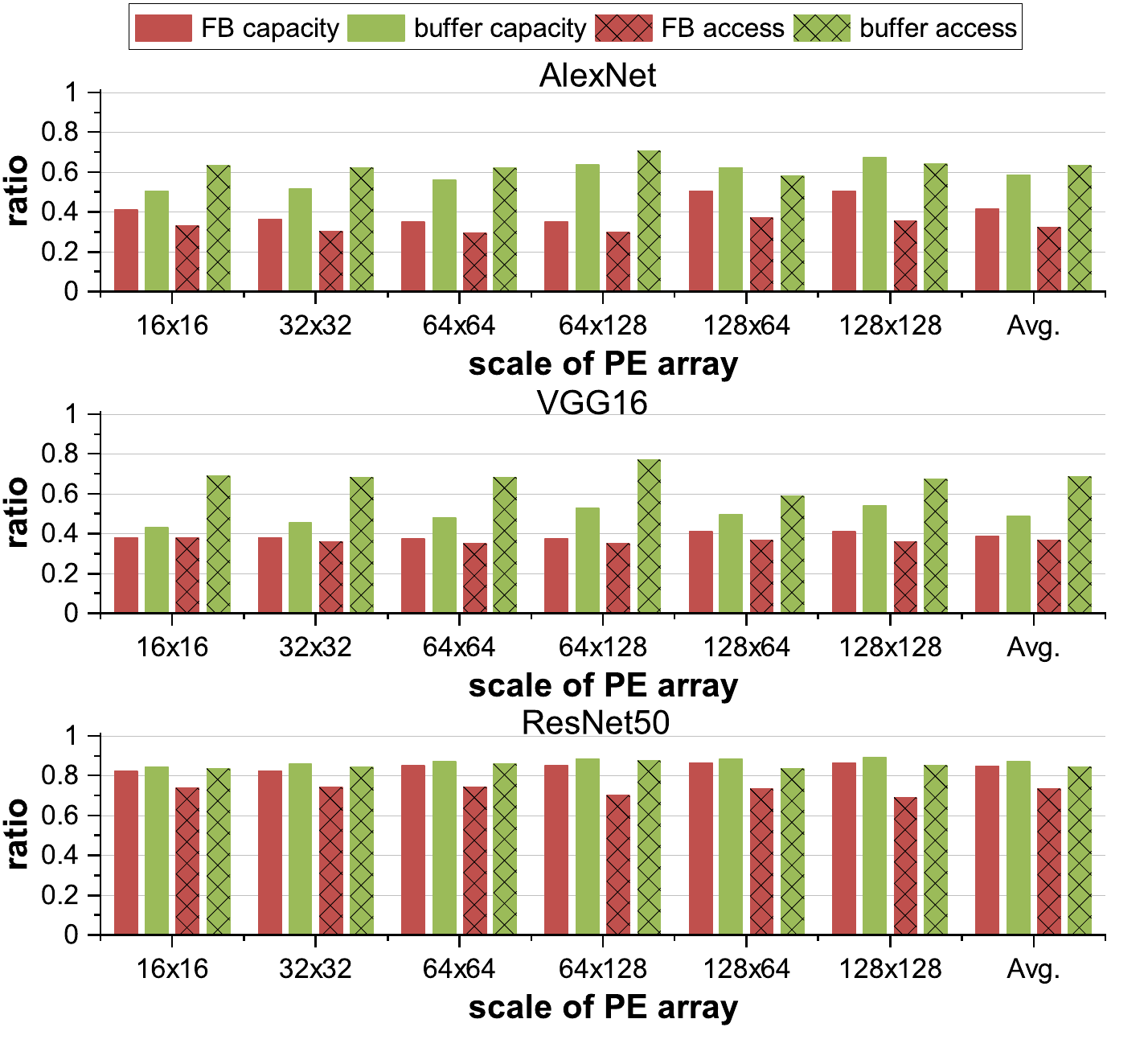}
    \vspace{-18pt}
    \caption{Analysis of reduction on buffer accessing and buffer capacity. }
    \label{fig14}
\end{figure}

Memory efficiency is demonstrated by evaluating the reduction of required buffer capacity and accessing when overlap reuse is introduced, which is exploited by CE array.
The average reduction ratios of all convolution layers in different CNN models are illustrated in \cref{fig14}.
A significant reduction on buffer capacity and accessing is achieved on AlexNet and VGG16 where $3 \times 3$ kernels are widely adopted. 
However, CE array cannot achieve such a significant reduction on ResNet50 due to the widely adopted $1 \times 1$ kernel.
Experimental results also show that \SSQ Engine with larger scale of PE array could obtain a slightly higher reduction both on buffer capacity and buffer accessing, which is reasonable because the inner data transmission between CEs are easier broken in smaller scale of PE array.
The following evaluation will show that such a reduction on memory accessing could largely improve its on-chip energy efficiency.

\subsection{Speed, Energy and Area Efficiency} \label{sec6_5}

\begin{figure}[t]
  \centering
  \includegraphics[width=\columnwidth]{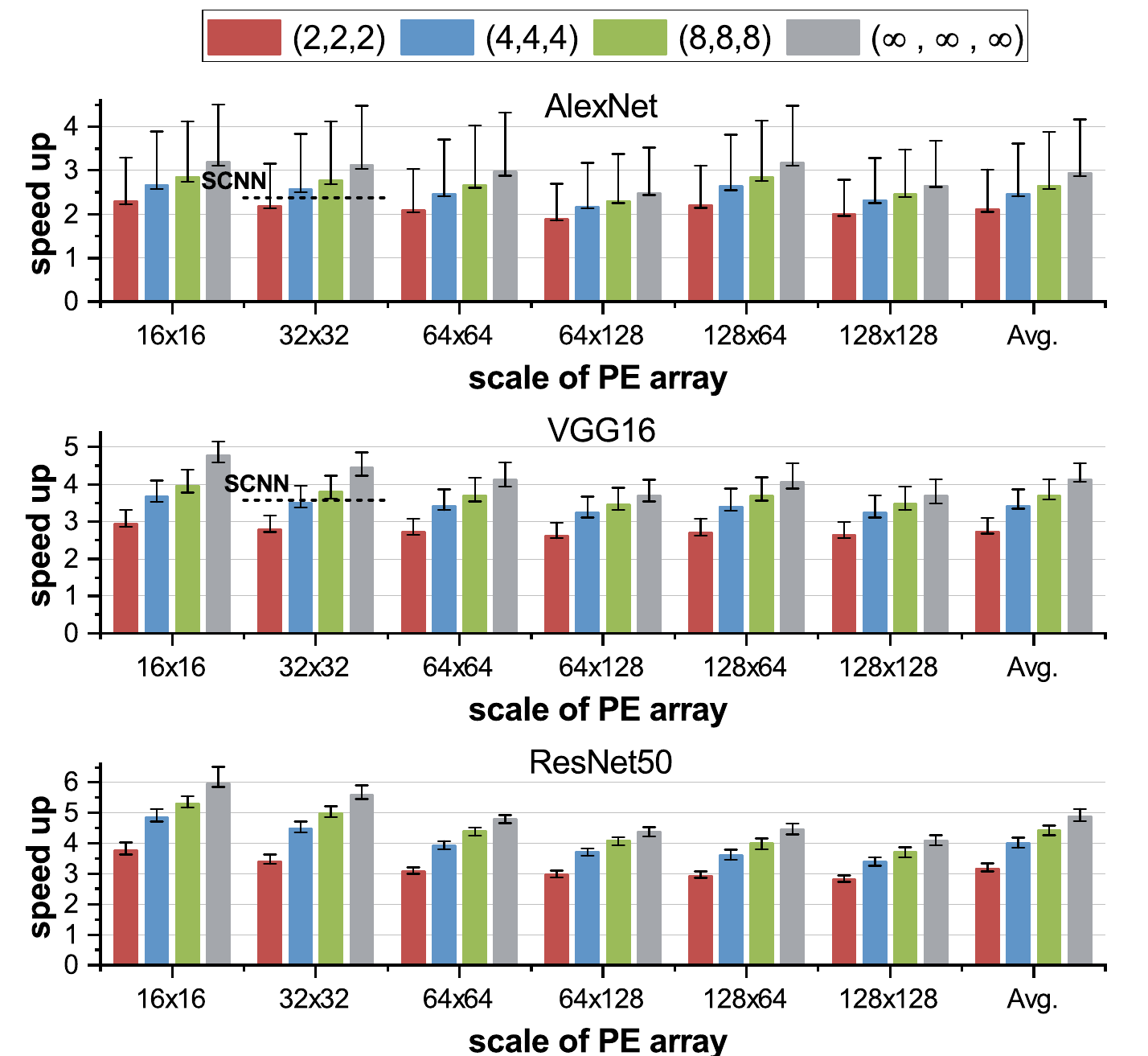}
  \caption{Speedups of \SSQ Engine with different scale of PE array and different FIFOs depth, when compared with na\"ive design across various CNN models. The histograms represent the speedups with average feature sparsity. The upper bound and lower bound for each histogram represents the speedup with maximum feature sparsity and minimum feature sparsity, respectively.}
  \label{fig15}
\end{figure}

\begin{figure}[t]
  \centering
  \includegraphics[width=\columnwidth]{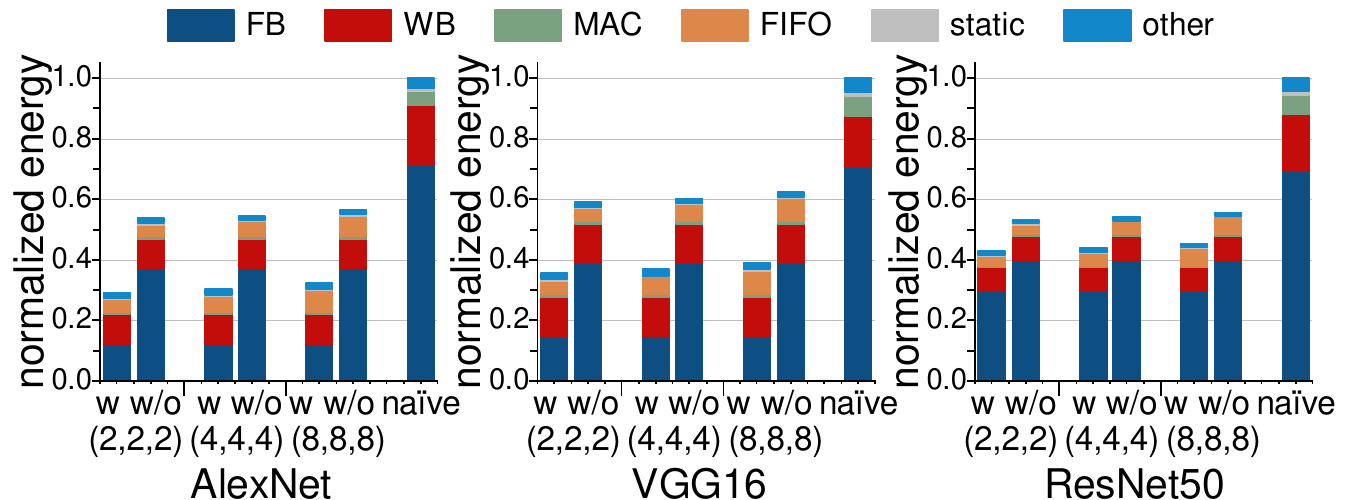}
  \caption{On-chip energy breakdown of \SSQ Engine with $16 \times 16$ PE array across various CNN models, where \texttt{w} means \SSQ Engine with CE array, \texttt{w/o} means \SSQ Engine without CE array.}
  \label{powerBreakdown}
\end{figure}

\begin{figure}[t]
  \centering
  \includegraphics[width=\columnwidth]{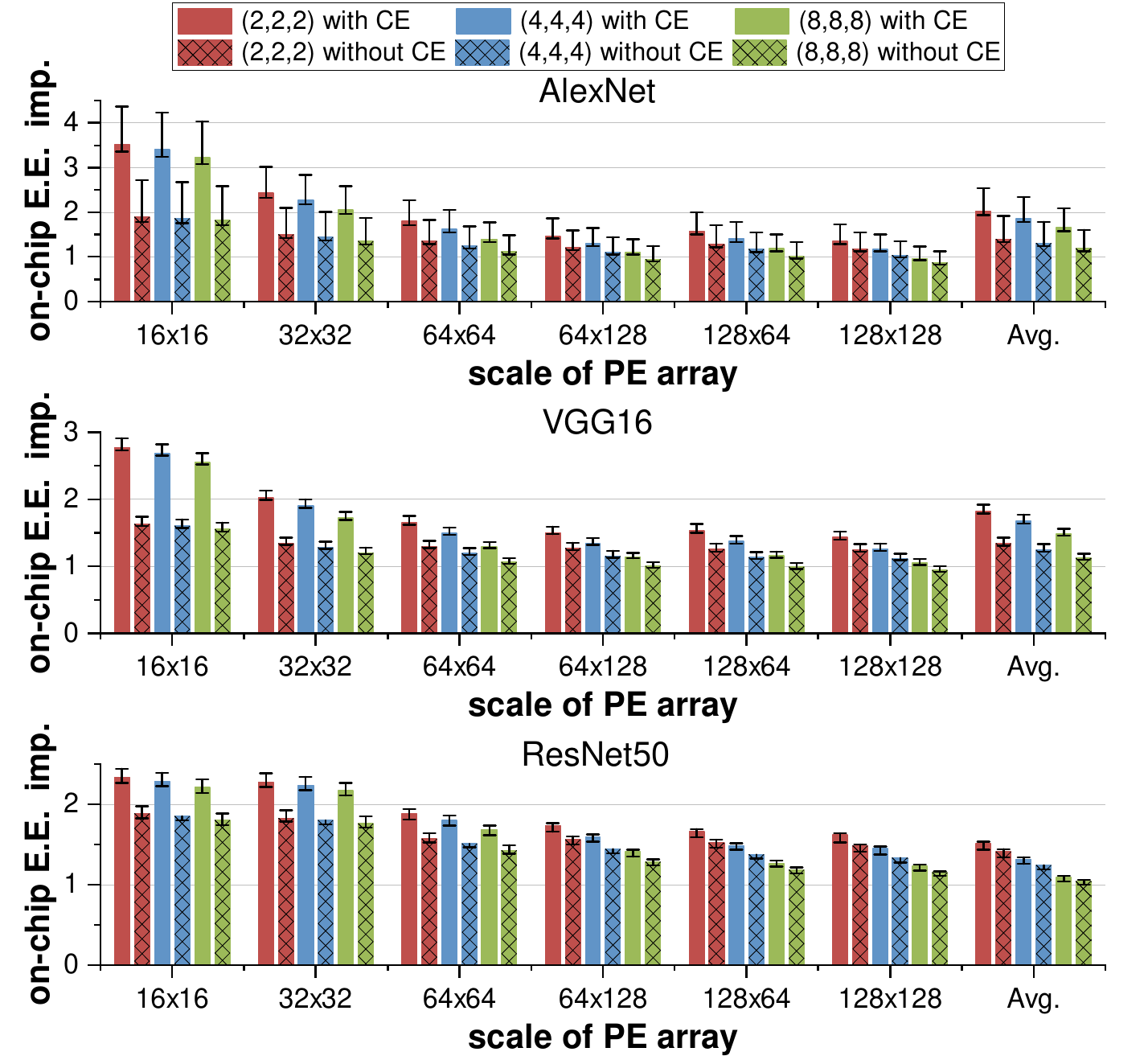}
  \caption{On-chip energy efficiency improvement (E.E. imp.) of \SSQ Engine with different scale of PE array and different FIFOs depth across various CNN models.}
  \label{energyEff}
\end{figure}

\begin{figure}[t]
  \centering
  \includegraphics[width=\columnwidth]{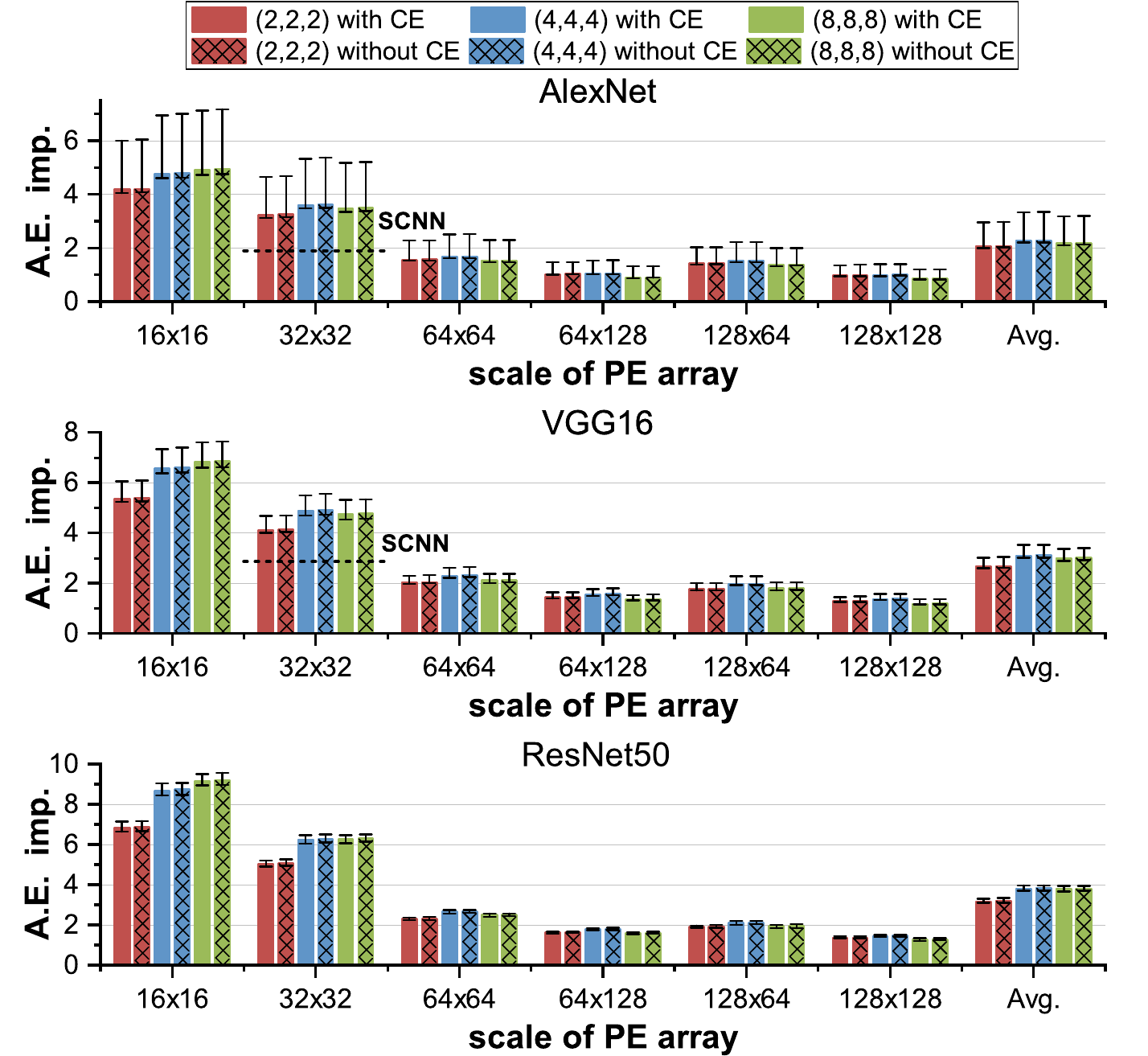}
  \caption{Area efficiency improvement (A.E. imp.) of \SSQ Engine with different scale of PE array and different FIFOs depth across various CNN models.}
  \label{areaEff}
\end{figure}

At last, actual sparse CNN models are evaluated on \SSQ Engine with different scales of PE array to identify the speedups and energy/area efficiency improvements.

\cref{fig15} illustrates the speedups of \SSQ Engine with different scale of PE array compared with na\"ive design. \SSQ Engine with larger scale of PE array will degrade the speedups because the dataflow usually becomes discontinuous after the DS procedure (as demonstrated in \cref{DSprocess}) and have to be accumulated PE by PE.
Compared with VGG16 and ResNet50, AlexNet has a larger upper bound of speedups than average speedups because it has a larger variance in feature density distribution according to the statistics in \cref{fig2}.
In summary, \SSQ Engine achieves about \textbf{3.2$\times$ speedups} on average versus the na\"ive design for all configurations and models. The largest speedup ($3.6\times$) is achieved when the FIFOs depth is $(8,8,8)$.
Such a speedup has almost reached the upper bound which is obtained with the FIFOs depth of $(\infty,\infty,\infty)$.
The main bottleneck left in this case is the frequency of DS component as evaluated in \cref{fig11}.

The energy consumption of DRAM access can be simply reduced by utilizing compressed dataflow, however, such an improvement is not very relevant to our proposed architectures.
Therefore, without considering the energy consumed by DRAM, the on-chip energy breakdown of \SSQ Engine with $16 \times 16$ PE array across various CNN models is illustrated in \cref{powerBreakdown}.
It shows that the utilized CE array could significantly reduce the energy consumption, and the reduction on energy consumption mainly comes from MAC and SRAM (FBs and WBs).
A large part of computations involved in MACs have been skipped which reduces the energy consumption significantly.
The energy consumption of SRAM access is reduced by utilizing the compressed dataflow while the energy consumption of FB could be further reduced by CE array.
It can be seen that the energy reduction achieved by these two aspects is much larger than the overhead introduced by FIFOs and therefore leads to better on-chip energy efficiency.
\cref{energyEff} further evaluates the benefits on energy consumption brought by \SSQ Engine in different scales of PE array and different FIFOs depth across various CNN models when compared with na\"ive design.
Actually, \SSQ Engine with CE array could achieve about $1.8\times$ improvement on average in energy efficiency than na\"ive design, and such an improvement has a good scalability as the PE array scales up. 
CE array contributes about $1.3\times$ to the improvement and it is more significant for smaller PE array because a large proportion of energy is consumed by FBs access.
The largest improvement ($1.9\times$) is achieved when FIFOs depth is set as $(2,2,2)$.
If further taking DRAM into consideration, \SSQ Engine could achieve about \textbf{3.0$\times$ improvement} of on-chip \textbf{energy efficiency} on average for different configurations and models.

The area efficiency improvement brought by \SSQ Engine in different scale is shown in \cref{areaEff}. 
The introduced CE array brings little impact on chip area since a small CE array is only enough. The working manner of PE array as illustrated in \cref{CEdataflow} shows that it would not introduce additional latency.
\SSQ Engine achieves significant area efficiency improvement especially for the case with smaller PE array in the same reason as analyzed in \cref{sec6_2}.
Consistent with the evaluation illustrated in \cref{fig14}, \cref{areaEff} also shows \SSQ Engine can achieve better area efficiency compared to SCNN on actual sparse networks.
However, as PE array scales up, the proportion of area occupied by PE array is gradually increased which diminishes the benefit brought by smaller SRAM and makes the overhead introduced by FIFOs more significant.
Therefore, \SSQ Engine achieves about $2.9\times$ area efficiency improvement on average for different PE array scales but only achieves about $1.2\times$ at the scale of $128\times128$.
As a tradeoff between speedup and area efficiency, the largest area efficiency ($3.1\times$) can be achieved when FIFOs depth of is set as $(4,4,4)$.

As a conclusion of the above evaluations, with different FIFOs depth, the average \textbf{speedup} (across different scales and networks) achieved by \SSQ Engine ranges from \textbf{2.7$\times$} to \textbf{3.6$\times$}, on-chip \textbf{energy efficiency} improvement ranges from \textbf{1.5$\times$} to \textbf{1.9$\times$}, \textbf{area efficiency} improvement ranges from \textbf{2.6$\times$} to \textbf{3.1$\times$}.
Therefore, it can be inferred that the performance would roughly fall into these intervals with various FIFOs depth, which shows the robustness of our approach.

Since SCNN and SparTen only report their improvement on speedup and energy efficiency versus their corresponding versions for processing dense networks, we can only provide a rough comparison here as presented in \cref{tabCompare}.
The scale of \SSQ Engine is set to $32\times32$ since it has the same number of multipliers as SCNN and SparTen. Meanwhile, only AlexNet and VGG16 are considered since they are evaluated for all designs.
With different configurations, \SSQ Engine could achieve \textbf{3.29$\times$} speedup or \textbf{2.70$\times$} energy efficiency improvement compared with the na\"ive design while SCNN could achieve up to 2.94$\times$ speedup and 2.21$\times$ energy efficiency improvement compared with its dense version. Although \SSQ Engine's speedup is not as good as that of SparTen ($5.60\times$), \SSQ Engine's energy efficiency improvement is much higher than SparTen's.
\cref{tabCompare} shows that our design can deploy the sparsity more efficiently and outperform SCNN under most of configurations, and is more energy efficient than SparTen. It can be also seen that our work is a kind of \textbf{lightweight} design from the area breakdown. It is obtained by utilizing the sparsity, e.g. dynamic selection and output-stationary dataflow. Our design is much more hardware friendly without introducing components with large overhead, compared with SCNN (involved scatter network and accumulator buffers) and SparTen (involved prefix-sum circuit and permute network).

\begin{table}[ht]
  \centering
  
  \caption{ Comparison of resources, area breakdown ($mm^2$), speedup, Energy Efficiency (E.E.) and Area Efficiency (A.E.) improvement (imp.) among \SSQ Engine, Na\"ive design, SCNN and SparTen. $*$ means $1.4\times$ and $0.5\times$ is E.E. imp. for partially memory and computation respectively.}
  
  \label{tabCompare}
  \resizebox{\linewidth}{!}{
    \renewcommand\arraystretch{1.5}
    \begin{tabular}{|c|c|c|c|c|c||c||c|}
    \hline
    \multicolumn{2}{|c|}{}           & \multicolumn{3}{c|}{\begin{tabular}[c]{@{}c@{}}\SSQ Engine  ($32\times32$)\end{tabular}} & \multirow{2}{*}{\begin{tabular}[c]{@{}c@{}}Na\"ive\\ ($32\times32$)\end{tabular}} & \multirow{2}{*}{SCNN} & \multirow{2}{*}{SparTen} \\ \cline{1-5}
    \multirow{3}{*}{ \begin{sideways}FIFO \end{sideways} }   & Depth  & 2 & 4  & 8   &  &   &    \\ \cline{2-8} 
      & Cap. & 12KB & 22KB & 32KB & -  & 32KB & 31KB   \\ \cline{2-8} 
      & Area  & 0.43   & 0.56    & 0.81   & -   & 0.26   & 3.20  \\ \hline
      \multirow{2}{*}{\begin{sideways}MULs\end{sideways}} & Num. & \multicolumn{4}{c||}{1024} & 1024 & 1024 \\ \cline{2-8} & Area & \multicolumn{4}{c||}{0.12} & 0.51 & 17.8 \\ \hline
      \multirow{2}{*}{\begin{sideways}SRAM\end{sideways}} & Cap. & \multicolumn{3}{c|}{1MB} & 2MB & 1MB & - \\ \cline{2-8} 
      & Area & \multicolumn{3}{c|}{1.44} & 2.89 & 1.98 & - \\ \hline
      \multicolumn{2}{|c|}{Total Area} & 2.03 & 2.15 & 2.39 & 3.04 & 7.9 & 24.5 \\ \hline
       \multicolumn{2}{|c|}{Technology} & \multicolumn{4}{c||}{14nm} & 16nm & 45nm \\ \hline
      \multicolumn{2}{|c|}{Speedup} & 2.49$\times$ & 3.05$\times$ & \textbf{3.29$\times$} & 1$\times$  & 2.94$\times$ & 5.60$\times$ \\ \hline
      \multicolumn{2}{|c|}{E.E. imp.}  & \textbf{2.70$\times$} & 2.66$\times$ & 2.59$\times$ & 1$\times$ & 2.21$\times$ & 1.4$\times$/0.5$\times^{*}$ \\ \hline
      \multicolumn{2}{|c|}{A.E. imp.}  & 3.67$\times$ & \textbf{4.23$\times$} & 4.11$\times$ & 1$\times$ & 2.20$\times$ & - \\ \hline
    \end{tabular}
}
\end{table}

\section{Related Works}\label{sec7}

Before the beginning of the era of machine learning, several systolic architecture-based accelerators were proposed for different algorithms \cite{6-asgari2014systolic, 7-hwang1989systolic}.
After the era of deep learning begins by \cite{9-krizhevsky2012imagenet}, many accelerators have been proposed for deep neural networks.
ShiDianNao \cite{8-du2015shidiannao} utilized a systolic array with kernel elements broadcast for convolutions.
The work in \cite{12-aydonat2017opencl} implemented AlexNet \cite{9-krizhevsky2012imagenet} with a 1-D systolic array.
UCLA \cite{10-wei2017automated} proposed an end-to-end automation flow from high level C code to a 2-D systolic array-based FPGA implementation.
Google's TPU \cite{11-jouppi2017datacenter} implemented an unprecedented $256 \times 256$ systolic architecture and achieved a great success in industry.
The work in \cite{kung2018packing} deployed the feature sparsity on a 2-D systolic array through column combining but the network could be only trained by their own pruning method.
Sparse-TPU \cite{he2020sparse} was similar to \cite{kung2018packing} and could exploit the feature sparsity as well as the weight sparsity, but its dataflow was not optimized for DNN applications.
Although these works have taken many explorations on some aspects, none of them could provide a fully supporting for sparse neural networks.

Eyeriss \cite{16-24-chen2017eyeriss} only optimized energy efficiency for memory access with the consideration of feature sparsity and was enhanced in \cite{chen2019eyeriss} to process sparse and compact networks.
Furthermore, Cnvlutin \cite{21-albericio2016cnvlutin} only utilized feature sparsity and Cambricon-X \cite{15-20-zhang2016cambricon} only considered weight sparsity. As somewhat an extension of Eyeriss, ZeNA \cite{kim2017zena} skipped all of the unnecessary computations but did not optimize the memory access.
SCNN \cite{17-parashar2017scnn} indeed fully exploited the sparsity with general pattern in both feature and weight, but required additional output coordinates transformation.
Cambricon-S \cite{22-zidong2018cambricon-s} also fully deployed the feature and weight sparsity but was only applicable for coarse-grained sparsity pattern.
DUET \cite{liu2020duet} could utilize both input and output feature sparsity through output speculation, but cannot support weight sparsity.
SparTen \cite{gondimalla2019sparten} could support sparse vector-vector multiplication and improve the hardware utilization, but its energy efficiency was significantly degraded due to the existed additional computation resources.
Eyeriss-V2 \cite{chen2019eyeriss} was also capable of processing compressed sparse data of both weight and feature, but its NoC consumed a large number of routers and the data transmission or dispatching mechanism was much complicated, which was a burden on performance and energy consumption.
In summary, none of these works could fully utilize the fine-grained sparsity both in feature and weight.

In regard to the mixed-precision quantization, Stripes \cite{13-judd2016stripes} only supported one kind of precision for each layer and TPU \cite{11-jouppi2017datacenter} could support $8$/$16$-bit processing in a very coarse granularity.
The work in \cite{14-park2018energy} introduced additional datapath for higher precision processing rather than reusing the computation logic resources, which significantly degraded the performance for processing higher precision cases.

\section{Conclusions}\label{sec8}

In this paper, \SSQ Engine is proposed as a novel systolic architecture for sparse CNNs acceleration. By allowing each PE to select the aligned weight-feature pairs from the passing cross compressed dataflow dynamically, \SSQ Engine solves the contradiction between the regularity of data transmission path and the irregularity of the sparsity. Consequently, full exploitation of the sparsity is achieved without restriction on sparse patterns. Furthermore, a CE array is exploited for efficient data reuse in systolic array as well as the extension of PEs for supporting fine-grained mixed-precision processing. \SSQ Engine is evaluated on ImageNet with real sparse CNN models. The experimental results have shown that it could achieve about $3.2\times$ speedup and about $3.0\times$ energy-efficiency improvement compared with the na\"ive systolic array.

\ifCLASSOPTIONcaptionsoff
  \newpage
\fi
{
\footnotesize 
\bibliographystyle{IEEEtran}
\bibliography{main.bib}
}
\vspace{-12mm}

\begin{IEEEbiography}[{\includegraphics[width=1in,height=1.25in,clip,keepaspectratio]{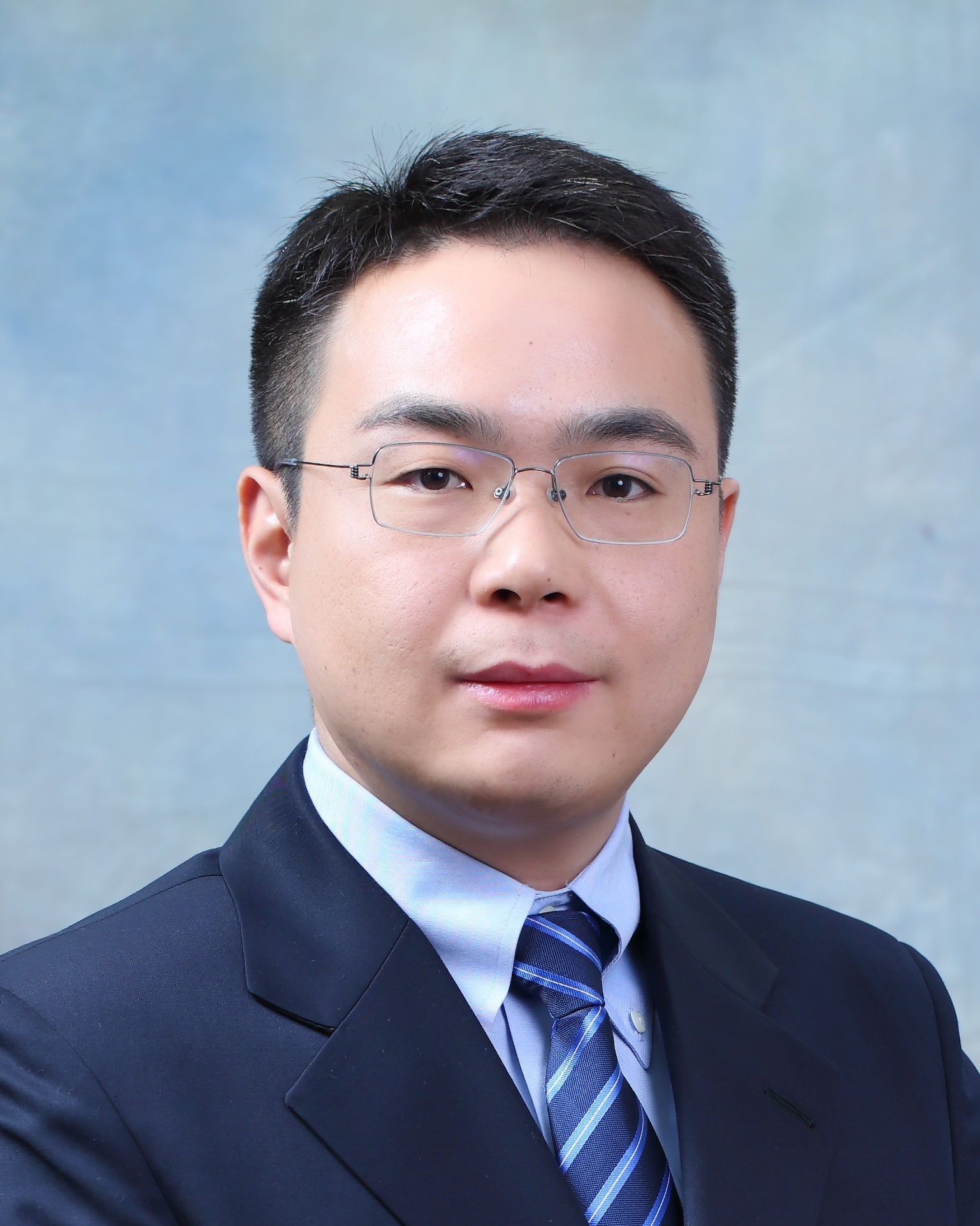}}]{Jianlei Yang}

(S'11-M'14-SM'20) received the B.S. degree in microelectronics from Xidian University, Xi'an, China, in 2009, and the Ph.D. degree in computer science and technology from Tsinghua University, Beijing, China, in 2014.

He is currently an Associate Professor in Beihang University, Beijing, China, with the School of Computer Science and Engineering. From 2014 to 2016, he was a post-doctoral researcher with the Department of ECE, University of Pittsburgh, Pennsylvania, United States.
His current research interests include computer architectures and neuromorphic computing systems.

Dr. Yang was the recipient of the First/Second place on ACM TAU Power Grid Simulation Contest in 2011/2012. He was a recipient of IEEE ICCD Best Paper Award in 2013, ACM GLSVLSI Best Paper Nomination in 2015, IEEE ICESS Best Paper Award in 2017, ACM SIGKDD Best Student Paper Award in 2020.

\end{IEEEbiography}
\vspace{-12mm}
\begin{IEEEbiography}[{\includegraphics[width=1in,height=1.25in,clip,keepaspectratio]{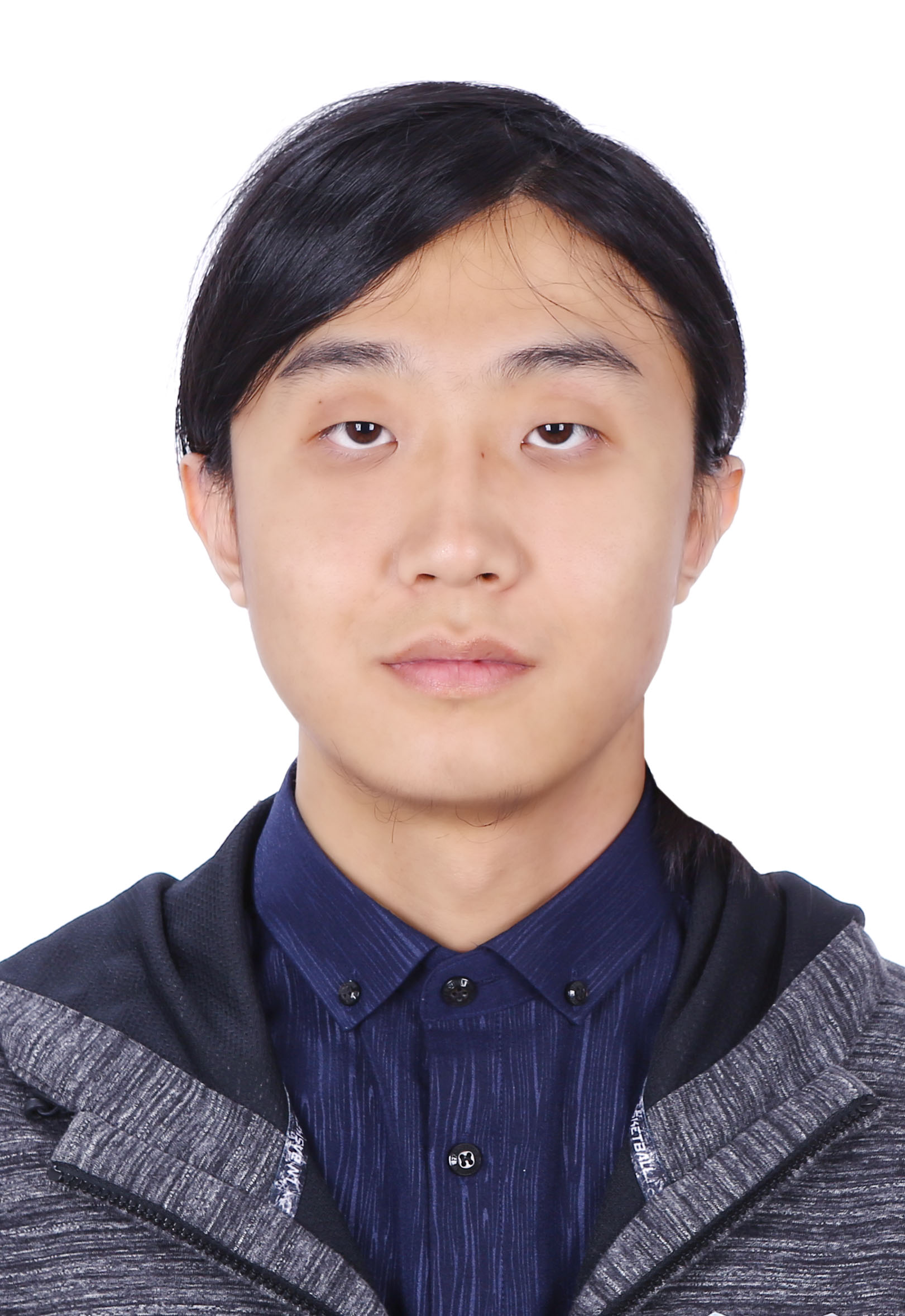}}]{Wenzhi Fu}

received the B.S. in computer science from Beihang University, Beijing, China, in 2019. He is currently a PhD student in School of Informatics, University of Edinburgh, Edinburgh, UK. His research interests include computing architectures, database theory and systems.

\end{IEEEbiography}
\vspace{-12mm}
\begin{IEEEbiography}[{\includegraphics[width=1in,height=1.25in,clip,keepaspectratio]{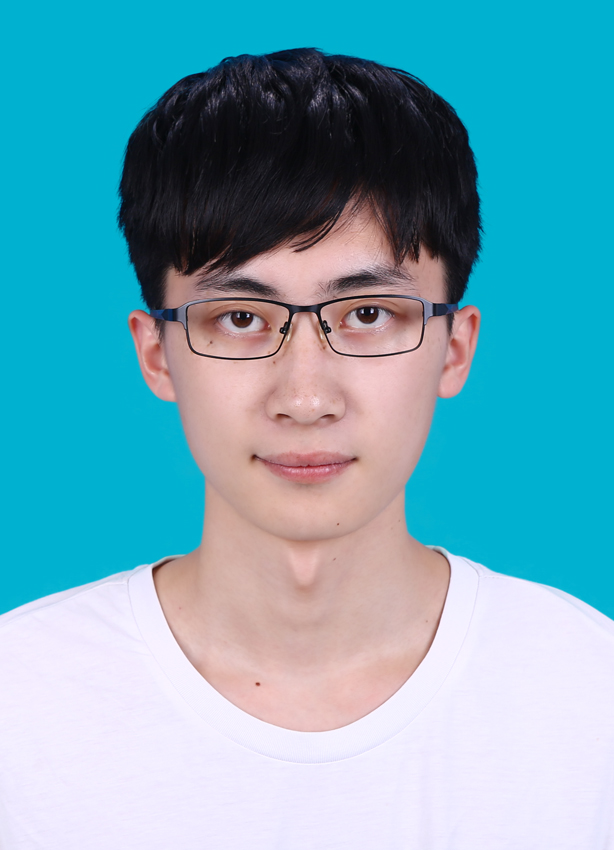}}]{Xingzhou Cheng}

received the B.S. in computer science from Beihang University, Beijing, China, in 2019. He is currently a graduate student in School of Computer Science and Engineering, Beihang University, Beijing, China. His research interests include computing architectures for deep learning and machine vision.

\end{IEEEbiography}
\vspace{-12mm}
\begin{IEEEbiography}[{\includegraphics[width=1in,height=1.25in,clip,keepaspectratio]{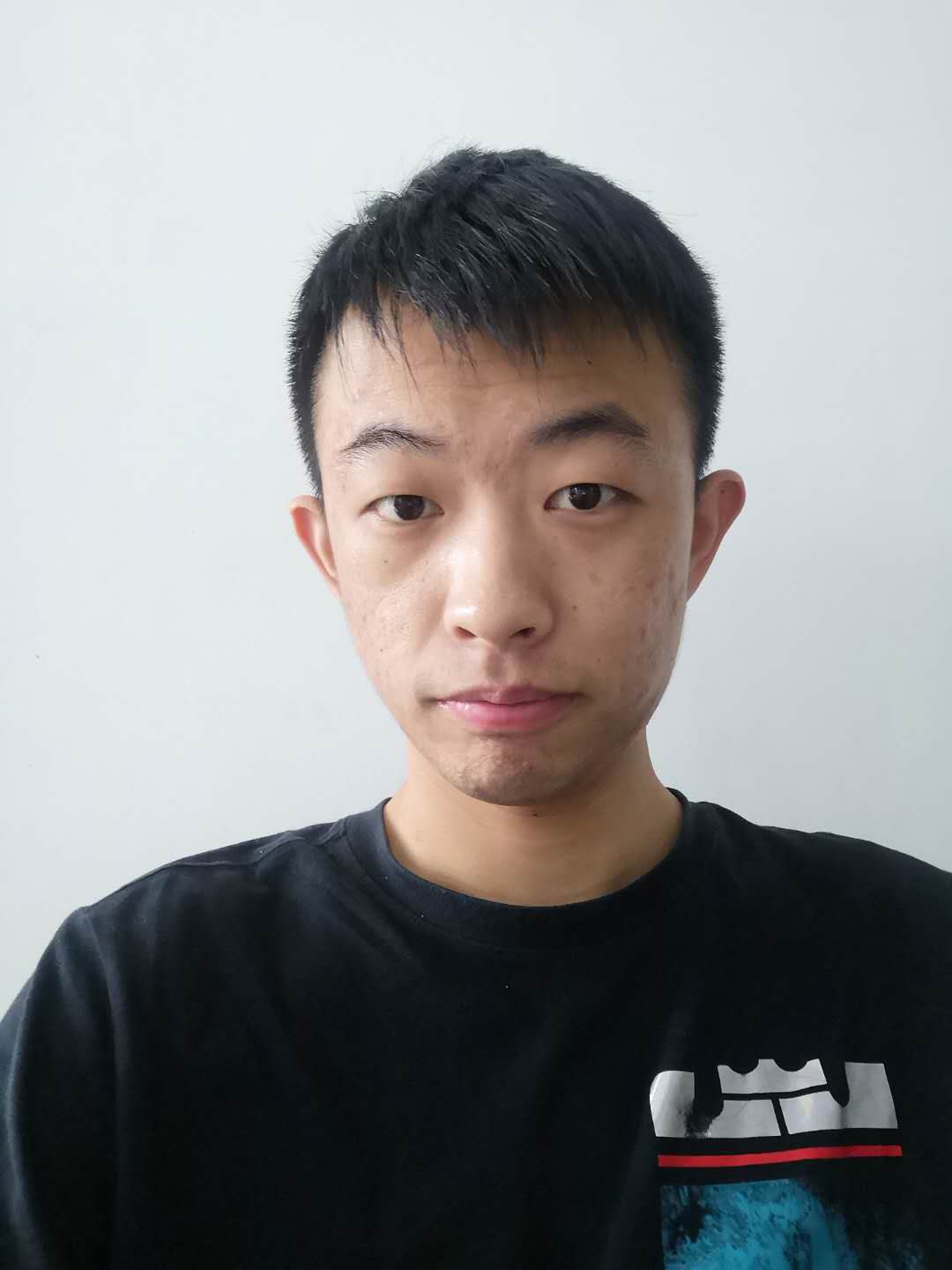}}]{Xucheng Ye}

received the B.S. in computer science from Beihang University, Beijing, China, in 2018. He is currently a graduate student in School of Computer Science and Engineering, Beihang University, Beijing, China. His research interests include deep learning algorithms and systems.

\end{IEEEbiography}
\vspace{-12mm}
\begin{IEEEbiography}[{\includegraphics[width=1in,height=1.25in,clip,keepaspectratio]{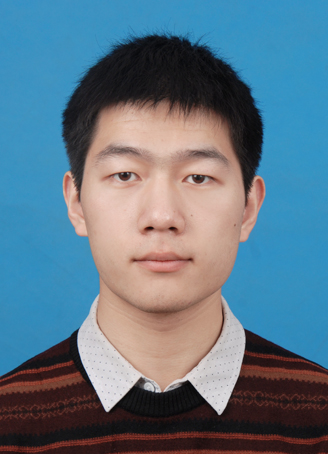}}]{Pengcheng Dai}

(S'18) received the B.S. and M.S. degrees in electronic engineering from Beihang University, Beijing, China, in 2017 and 2020, respectively. He is currently a software development engineer in ByteDance Ltd. Inc., Beijing, China. His research interests include computing architectures for deep learning and machine vision.

\end{IEEEbiography}
\vspace{-12mm}
\begin{IEEEbiography}[{\includegraphics[width=1in,height=1.25in,clip,keepaspectratio]{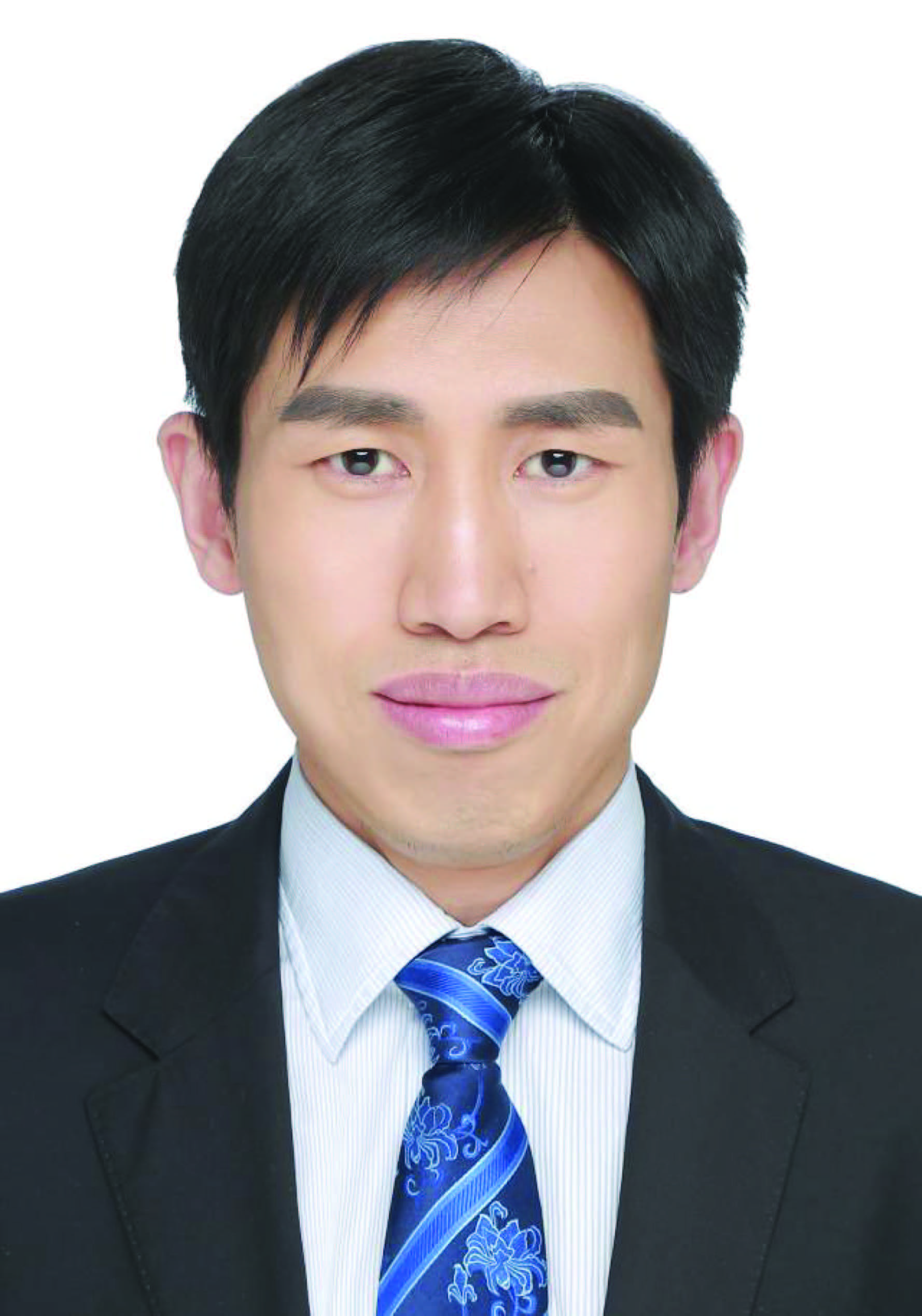}}]{Weisheng Zhao}

(M'06-SM'14-F'19) received the Ph.D. degree in physics from University of Paris Sud, Paris, France, in 2007. He is currently the Professor with the School of Microelectronics in Beihang University, Beijing, China.

In 2009, he joined the French National Research Center(CNRS), as a Tenured Research Scientist. Since 2014, he has been a Distinguished Professor with Beihang University, Beijing, China. He has published more than 200 scientific articles in leading journals and conferences, such as \textsc{Nature Electronics, Nature Communications, Advanced Materials, IEEE Transactions}, ISCA and DAC. His current research interests include the hybrid integration of nano-devices with CMOS circuit and new nonvolatile memory (40-nm technology node and below) like MRAM circuit and architecture design.

He is currently the Editor-In-Chief for the \textsc{IEEE Transactions on Circuits and Systems I: Regular Paper}. He is an IEEE Fellow.

\end{IEEEbiography}

\end{document}